\pdfoutput=1
\documentclass[a4paper,11pt]{article}
\usepackage{jheppub} 
\usepackage{comment}
\usepackage{amsmath,amssymb,amsfonts,mathtools}
\usepackage[toc,page]{appendix}
\usepackage{hyperref}
\usepackage{physics}
\usepackage{float}
\usepackage{slashed}
\usepackage[dvipsnames]{xcolor}
\usepackage{tikz}
\usetikzlibrary{decorations.markings, arrows.meta, decorations.pathmorphing}

\newcommand{\hmntt}[0]{h_{\mu \nu}^{TT}}
\newcommand{\TTproj}[0]{\varepsilon_\mn^{TT,\ab}}

\newcommand{\sfU}[0]{\mathsf{U}}
\newcommand{\sfP}[0]{\mathsf{P}}
\newcommand{\sfA}[0]{\mathsf{A}}

\newcommand{\tx}[1]{ \ \text{#1} \ }
\newcommand{\lara}[1]{\left\langle #1 \right\rangle}

\newcommand{\ab}[0]{{\alpha \beta}}

\newcommand{\mn}[0]{{\mu \nu}}

\newcommand{\MN}[0]{{M N}}

\renewcommand{\ij}[0]{{ij}}

\newcommand{\pd}[0]{\partial}

\newcommand{\mt}[1]{\textrm{\scriptsize #1}}

\newcommand{\A}[0]{\mathcal{A}}

\newcommand{\abe}[0]{{ab}}

\newcommand{\C}[0]{\mathcal{C}}
\newcommand{\D}[0]{\mathcal{D}}

\newcommand{\F}[0]{\mathcal{F}}

\newcommand{\I}[0]{\mathcal{I}}

\renewcommand{\O}[0]{\mathcal{O}}
\renewcommand{\P}[0]{\mathcal{P}}

\newcommand{\R}[0]{\mathcal{R}}
\renewcommand{\S}[0]{\mathcal{S}}

\newcommand{\liegroup}[3]{\text{#1}\qty(#2)^{#3}}

\renewcommand{\dd}{\mathrm{d}}
\def\Nc{N_\mt{c}}
\def\Nf{N_\mt{f}}

\newcommand{\be}{\begin{equation}}
\newcommand{\ee}{\end{equation}}
\newcommand{\bea}{\begin{eqnarray}}
\newcommand{\eea}{\end{eqnarray}}

\newcommand{\alphasymbol}[0]{\Omega}

\makeatletter
\def\@fpheader{\relax}
\makeatother

\preprint{\text{HIP-2026-25/TH}}

\title{Gravitational form factors of pions in V-QCD}

\author[a]{Antti Hippel\"ainen,}
\author[a]{Niko Jokela,} 
\author[b]{and Matti J\"arvinen}
\affiliation[a]{Department of Physics and Helsinki Institute of Physics, \\
P.O. Box 64, FI-00014, University of Helsinki, Finland}
\affiliation[b]{Institute of Theoretical Physics, Chinese Academy of Sciences, Beijing 100190, China}

\emailAdd{antti.hippelainen@helsinki.fi}
\emailAdd{niko.jokela@helsinki.fi}
\emailAdd{mattijarvinen@itp.ac.cn}

\abstract{We compute the gravitational form factors and mechanical properties of the pion in V-QCD, a holographic model of QCD with flavor backreaction and dynamical chiral symmetry breaking. The model parameters are taken from earlier fits to meson spectra and are not adjusted to fit form factor data. We extract the matrix element of the energy-momentum tensor in the pion state following the holographic dictionary and show that the momentum sum rule $\A(0)=1$ then holds automatically. The contributions to $\D(t)$ from two different diagrams are separately singular at $t=0$, but these singularities cancel each other exactly. At the physical pion mass, we obtain $\D(0)\approx -0.98$, close to the chiral-limit value $-1$ of the soft-pion theorem. We compute the pressure, shear force, and energy density distributions of the pion and estimate its mechanical and mass radii as 0.86~fm and 1.32~fm, respectively. We also compare our results with those from lattice QCD, dispersive analysis, Dyson--Schwinger methods, and an analysis of experimental pion-pair production data.}

\begin{document}
\maketitle
\flushbottom
\setcounter{page}{2}

\clearpage

\section{Introduction}
\label{sec:intro}

Gravitational form factors (GFFs) provide an important tool for probing the properties of hadrons. Defined through the matrix elements of the QCD energy-momentum tensor in a single-hadron state, they provide information about the distribution of energy, momentum, and internal stress of the hadron, complementing the structure probed by matrix elements of electromagnetic currents. For a spin-zero hadron, such as the pion, there are two independent GFFs $\A(t)$ and $\D(t)$, where $t$ denotes the invariant momentum transfer. The momentum sum rule fixes $\A(0)=1$, while $\D(t)$ determines the pressure and shear distributions associated with the internal forces. For a review, see~\cite{Polyakov:2018zvc}.

The pion is of particular interest because it is both a composite hadron and a pseudo-Nambu--Goldstone boson of spontaneously broken chiral symmetry: Its small mass reflects the weak explicit breaking of the chiral symmetry by the light quark masses on top of the spontaneous breaking of the symmetry, whereas its structure is dictated by strong QCD dynamics. Chiral symmetry also enters the gravitational response of the pion as the soft-pion theorem fixes $\D(0)=-1$ in the chiral limit~\cite{Novikov:1980fa,Voloshin:1980zf,Voloshin:1982eb,Leutwyler:1989tn, Polyakov:1999gs}.
This provides a nontrivial cross-check for computations in the limit of small pion mass. The mass corrections to $\D(0)$ and the momentum dependence of the form factors probe the nonperturbative nature of the pion dynamics.

Although GFFs cannot be measured directly, they can be extracted from experiments: They are related to moments of generalized parton distributions (GPDs)~\cite{Ji:1996nm} in hard scattering and generalized distribution amplitudes (GDAs)~\cite{Diehl:1998dk} in hadron-pair production, which can be accessed experimentally. Indeed, this connection has been used to extract the pion GFFs from Belle $\gamma^*\gamma\to\pi^0\pi^0$ data~\cite{Belle:2015oin} using GDAs~\cite{Kumano:2017lhr}. Upcoming experiments at the Electron-Ion Collider (EIC) provide strong motivation for further study. At the EIC, pion structure will be probed via the Sullivan process, where an electron scatters off a virtual pion emitted by a proton while a recoiling neutron tags the exchange~\cite{Chavez:2021koz}. Moreover, the near-threshold vector meson production has been proposed as a direct experimental probe of the pion GFFs~\cite{Hatta:2025ryj}.

Pion GFFs have been computed in lattice QCD~\cite{Hackett:2023nkr}, building on earlier determinations of quark and gluon contributions~\cite{Brommel:2005ee,Shanahan:2018pib,Pefkou:2021fni}, as well as via Dyson--Schwinger methods~\cite{Xu:2023izo} and dispersive analyses~\cite{Cao:2024zlf,Cao:2025dkv}. Beyond these approaches, several effective and phenomenological models have been used to extract the GFFs, including the Nambu--Jona-Lasinio model~\cite{Broniowski:2008hx,Freese:2021czn}, relativistic and light-front quark models~\cite{Krutov:2020ewr,Choi:2025rto}, the instanton liquid model~\cite{Liu:2024jno,Liu:2024vkj}, and meson dominance~\cite{Broniowski:2024oyk}. Pion GFFs have also been explored in dilaton effective theory~\cite{Zwicky:2023fay}. Motivated by this framework, scalar-pole parametrizations incorporating a $\sigma/f_0(500)$ resonance and a smooth background have been used to describe lattice results for the total and gluon $\D$ form factors~\cite{Stegeman:2025sca,Stegeman:2025tdl}.

Holographic models provide a complementary approach for analyzing GFFs. An early AdS/QCD calculation determined $\A(t)$ from the tensor response but left $\D(t)$ undetermined~\cite{Abidin:2008hn}. This study was later extended to other AdS/QCD models~\cite{Deng:2026suw} and to include flavor effects, giving different predictions for pions and kaons~\cite{Liu:2025vfe}. Recent works computing both form factors in the chiral limit in the Sakai--Sugimoto model can be found in~\cite{Fujii:2024rqd,Fujii:2025eug}.

Our recent calculation of proton GFFs in improved holographic QCD (IHQCD)~\cite{Gursoy:2007cb,Gursoy:2007er}~ combined the tensor and scalar gravitational fluctuations with five-dimensional Dirac fields representing the proton~\cite{Hippelainen:2026izm} (see also~\cite{Abidin:2009hr,Mamo:2019mka,Fujita:2022jus,Sugimoto:2025btn,Deng:2025fpq,Deng:2026gik}). In that calculation, the apparent pole in the gluon form factor $\D_g(t)$ at vanishing momentum transfer cancels analytically, leaving a finite $\D_g(0)$. In this work we extend this study to the pion in a model with dynamical flavor and chiral symmetry breaking. The spin-one mesons of the same model are treated in~\cite{Hippelainen:2026xko}. In contrast to the Sakai--Sugimoto calculations of Refs.~\cite{Fujii:2024rqd,Fujii:2025eug}, the quark mass is nonzero, and the flavor sector backreacts with gluon sector.

The holographic model which we use in the analysis of the pion GFFs is V-QCD, a bottom-up holographic model of QCD with a fully backreacted quark flavor sector on top of the gluon dynamics~\cite{Jarvinen:2011qe}. The glue sector of V-QCD is the IHQCD model, while the flavor sector is described by a Dirac--Born--Infeld action with chiral symmetry breaking described by the condensation of a scalar field dual to the quark bilinear operator, the `tachyon.' We use the potentials and parameters obtained from the meson spectrum fit of~\cite{Amorim:2021gat} at the physical pion mass, $m_\pi\approx 135$~MeV. We also consider $m_\pi\approx 170$~MeV to compare directly with the lattice calculation. None of the model parameters are adjusted by using GFF data. 

After fixing the model, we must identify the matrix element from the three-point function using holographic Lehmann--Symanzik--Zimmermann (LSZ) reduction~\cite{Hoyos:2019kzt,Hoyos:2020fjx,Hoyos:2022ptd}. After this we can identify the form factors from the gravitational three-point vertices. We find $\A(0)=1$ and, at the physical pion mass, $\D(0)\approx -0.984$. The latter is close to the chiral-limit soft-pion value $-1$ and to the one-loop chiral perturbation theory estimate $\D(0)=-0.97 \pm 0.01$ of the gravitational low-energy constants~\cite{Donoghue:1991qv, Hudson:2017xug}.
We compare our results for  $\A(t)$ and $\D(t)$ to lattice data and other theoretical results, finding good agreement, given that no parameters were fitted to the data. We also compute the pressure, shear force, and energy density distributions. At the physical pion mass, the mechanical and mass radii are found to be approximately $0.86$ fm and $1.32$ fm, respectively. 

The article is organized as follows. We discuss the holographic model, including the background geometry and its fluctuations, in Section~\ref{sec:holographicmodel}. We then analyze the pion GFFs in Section~\ref{sec:gravitationalformfactors}, showing how they arise from three-point correlators strictly following the standard holographic dictionary. We compute them numerically in our model and compare them with results from other approaches. We use these results to analyze the mechanical properties of the pion in Section~\ref{sec:mechanicalproperties}. Finally, we conclude in Section~\ref{sec:conclusions} with a discussion about possible future extensions of our work. The appendices contain additional technical details.

\section{Holographic model}\label{sec:holographicmodel}

We consider a five-dimensional holographic model, V-QCD~\cite{Jarvinen:2011qe}. The ``V'' in the name of the model refers to the fact that it originates from considering a noncritical string theory background in the Veneziano limit~\cite{Veneziano:1976wm}, where the numbers of colors $\Nc$ and flavors $\Nf$ are taken to infinity, and the ratio $x \equiv {\Nf}/{\Nc}$ is kept constant. However, in phenomenological applications one typically sets $\Nf$ and $\Nc$ directly to the values of physical QCD, with the understanding that the $1/N$-corrections may be captured through matching the model with data at finite values of these parameters. 
Here, we set $x = {2}/{3}$, corresponding to two active quark flavors and $\Nc = 3$.

In addition to the metric tensor, the model includes a dilaton $\Phi$ and a tachyon field $T$. The metric is dual to the energy-momentum tensor $T^\mn$, the dilaton $\Phi$ to $\Tr G_\mn G^\mn$, where $G_\mn$ is the field-strength tensor of gluons, the tachyon $T^{ij}$ to the quark bilinear $\bar{q}^i_R q^j_L$, and gauge fields $L_M^{ij}, R_M^{ij}$ to the left- and right-handed flavor currents. Our notation is such that capital Latin indices $M, N, \dots$ run over five dimensions, Greek $\mu, \nu,\dots$ over the coordinates of the four-dimensional boundary, and $i, j = 1, \ldots , \Nf$ are flavor indices. 

The gluon sector of the model is described through the IHQCD model~\cite{Gursoy:2007cb,Gursoy:2007er}.
On top of IHQCD, dynamical flavor is included in the model by a pair of space-filling $\text{D4}-\overline{\text{D4}}$ branes with an open-string tachyon~\cite{Bigazzi:2005md,Casero:2007ae,Bergman:2007pm,Dhar:2007bz,Dhar:2008um,Jokela:2009tk}, and the dynamics are described by a Dirac--Born--Infeld action~\cite{Leigh:1989jq}. The action reads~\cite{Jarvinen:2011qe}
\begin{equation}
    \label{eq:VQCDaction}
    \begin{split}
        S &= S_g + S_f \\
        S_g &= M^3 \Nc^2 \int \dd^5 x \sqrt{-g} \qty(\R  - \frac{4}{3}(\pd \Phi)^2 + V_g(\Phi))   \\
        S_f &= - \frac{1}{2} M^3 \Nc \Tr \int \dd^5 x \qty[V_f(\Phi, T^\dagger T)\sqrt{- \text{det} \boldsymbol{L}} + V_f(\Phi,  TT^\dagger)\sqrt{- \text{det} \boldsymbol{R}}] \ .
    \end{split}
\end{equation}
Here, $M$ is the Planck mass, $\Tr$ denotes the symmetrized trace over the (suppressed) flavor indices, $g$ is the determinant of the metric, and $\R$ is the Ricci scalar. The functions $V_f$ and $V_g$ are potentials (see Appendix~\ref{app:potentials}). The full action of V-QCD also contains the CP-odd sector with an axion term, but it does not contribute to the analysis in this work and is omitted. 

In the DBI action, the matrices $\boldsymbol{L}$, $\boldsymbol{R}$ are defined as 
\begin{equation}
\label{eq:gaugefieldterms}
    \begin{split}
        \boldsymbol{L}_\MN &= g_\MN  + w(\Phi) F_\MN^L + \frac{\kappa(\Phi)}{2}\qty[(\D_M T)^\dagger (\D_N T) + (\D_N T)^\dagger (\D_M T)]   \\
        \boldsymbol{R}_\MN &= g_\MN  + w(\Phi) F_\MN^R + \frac{\kappa(\Phi)}{2}\qty[(\D_M T) (\D_N T)^\dagger + (\D_N T) (\D_M T)^\dagger] \ ,
    \end{split}
\end{equation}
where we have added two new potentials, $w(\Phi)$ and $\kappa(\Phi)$. We have left out a possible $T$-dependence in both of these potentials. We have suppressed the flavor dependence of the gauge fields. 

The standard covariant derivative and non-Abelian field strength in the above are
\begin{equation}
    \begin{split}
        \D_M T &= \pd_M T  + i T L_M - i R_M T  \\
        F_\MN^L &= \pd_M L_N - \pd_N L_M - i \comm{L_M}{L_N}  \ ,
    \end{split}
\end{equation}
and analogously for the right-handed gauge field. We use the potential parametrizations and parameters given in~\cite{Amorim:2021gat}, fit to meson spectra~\cite{Arean:2012mq,Arean:2013tja}. The potentials are defined explicitly in Appendix~\ref{app:potentials}. Apart from the potentials, fully defining the model requires fixing the Planck mass $M$ in~\eqref{eq:VQCDaction} and the intrinsic scale $\Lambda_{\rm UV}$ of the holographic RG flow~\cite{Gursoy:2007cb,Gursoy:2007er,Jarvinen:2011qe}. We do not need the value of $M$ in this article, so we leave it unspecified (it could be determined, for example, by matching with the pion decay constant). The value of $\Lambda_{\rm UV}$ gives natural units for all dimensionful observables; see Appendix~\ref{app:potentials} for its numerical values and for more discussion.

\subsection{Gravity background}\label{subsec:gravitybackground}

At the level of the background, all fields are functions of the holographic coordinate $r$ only. For the metric, we use the Ansatz
\begin{equation}
    ds^2 = e^{2A(r)} \eta_\MN \dd x^M \dd x^N = e^{2A(r)}\qty(- \dd t^2 + \dd \mathbf{x}^2 + \dd r^2) \ ,
\end{equation}
where $M, N = 0,\dots,4$, and $\eta_\MN = \text{diag}(-1,1,1,1,1)$. The gauge fields have no background value, $L_M = R_M = 0,$ while the dilaton and tachyon are given as $\Phi\qty(x^M) = \Phi(r), T^{ij}\qty(x^M) = \tau(r) \delta^{ij}$. That is, the flavors are identical, and the aim of the construction is to model two light quarks having the same mass. 
The flavor action becomes
\begin{equation}
\label{eq:backgroundflavoraction}
    S_f = - M^3 \Nc \Nf \int \dd^5 x V_f(\Phi, \tau) \sqrt{-\det \qty(g_\MN + \kappa(\Phi) (\pd_M \tau)(\pd_N \tau))} \ .
\end{equation}

The holographic interpretation of $A(r)$ is the logarithm of the energy scale, since scalings of spacetime coordinates can be absorbed into shifts of $A(r)$. The geometry starts at $r = 0$, where the UV boundary is located, while it ends in an IR singularity at $r \to \infty$. Unless required for clarity, explicit dependence of fields on $r$ is suppressed throughout. After the action is fixed, the backgrounds depend on a single (dimensionless) parameter, which can be defined in terms of the IR asymptotics of the tachyon field~\cite{Jarvinen:2011qe} (see Appendices~\ref{app:potentials} and~\ref{subapp:solvingthebackground}) or equivalently the quark mass or the pion mass. Following the fit of~\cite{Amorim:2021gat}, we determine this parameter through the value of the pion mass.

The background is obtained by solving the combined set of Einstein equations and the equations of motion for $\Phi$ and $\tau$. The full set of background equations of motion reads~\cite{Jarvinen:2011qe} 
\begin{equation}
\label{eq:backgroundeoms}
    \begin{split}
        3\qty(A'' + (A')^2) &= - \frac{2}{3} \qty(\Phi')^2 + \frac{1}{2} e^{2A}V_g - \frac{1}{2} x V_f e^{2A} G  \\
        6 \qty(A')^2 &= \frac{2}{3}\qty(\Phi')^2 + \frac{1}{2}e^{2A}V_g - \frac{1}{2}x V_f \frac{e^{2A}}{G}  \\
        0 &= \Phi'' + 3 A' \Phi' + \frac{3}{8}e^{2A} \qty(\pdv[]{V_g}{\Phi} - x G \pdv[]{V_f}{\Phi} - e^{-2A} x V_f \pdv[]{\kappa}{\Phi} \frac{\qty(\tau')^2}{2G})   \\
        0 &= e^{5A} \pdv[]{V_f}{\tau} G 
        - \pd_r \qty(e^{3A} V_f \kappa \frac{\tau'}{G})  \ ,
    \end{split}
\end{equation}
where we have defined the ubiquitous factor 
\begin{equation}
\label{eq:Gdef}
    G(r) = \sqrt{1 + e^{-2A} \kappa (\tau')^2} \ .
\end{equation}
Throughout the paper, a prime denotes $\pd_r$ unless otherwise stated. 

Solving these equations numerically is challenging because the IR behavior of the tachyon makes the equations stiff. A brief description of the numerical method is given in Appendix~\ref{subapp:solvingthebackground}.

\subsection{Gravity fluctuations}\label{subsec:gravityfluctuations}

In this section we parametrize the metric, dilaton, and tachyon fluctuations, organize the scalar fluctuations into gauge-invariant combinations, and define the bulk-to-boundary propagators needed for the vertices in Section \ref{sec:gravitationalformfactors}.

We will expand the action with respect to both tensor and scalar fluctuations and must parametrize the full five-dimensional versions of the dynamical fields. The dilaton and tachyon are 
\begin{equation}
    \Phi\qty(x^M) = \Phi(r) + \chi\qty(x^M) \ , \qquad T\qty(x^M) = \tau(r) + s\qty(x^M) 
    \ ,
\end{equation}
where $\chi$ and $s$ are the scalar infinitesimal fluctuations. The fluctuated metric is set up as \cite{Arean:2013tja, Kiritsis:2006ua} 
\begin{equation}
\label{eq:fluctuatedmetric}
\dd s^2 = e^{2A(r)}\left[(1+2 \phi)\dd r^2  +2 \partial_\mu W\dd r \dd x^\mu  + \qty((1+2 \psi)\eta_{\mu\nu} + 2 \partial_\mu \partial_\nu E  + 2h_{\mu\nu}^{TT}) \dd x^\mu \dd x^\nu\right] \ ,
\end{equation}
where the fields $\phi, W, \psi, E$, and $\hmntt$ are all infinitesimal and depend on $x^M$, and $\eta_{\mu\nu}$ is the Minkowski metric in the mostly plus convention. The mode $\hmntt$ is transverse and traceless, $\pd^\mu \hmntt = \eta^\mn \hmntt = 0$. We could have also included a transverse vector and a transverse axial field in the fluctuation of the metric, but these terms do not contribute to the final result, since the possible gauge-invariant vector mode does not propagate. 

Not all of the fluctuation fields introduced above are physical, since some of them can be removed by diffeomorphism invariance. Scalar infinitesimal diffeomorphism transformations
$r \mapsto r + \xi_r$ and $x^\mu \mapsto x^\mu + \pd^\mu \xi$ are parametrized in terms of two functions, $\xi$ and $\xi_r$. Under this transformation, the line element changes, and the fluctuations of fields must compensate for this change. The scalar fluctuations  then transform as
\begin{equation}
    \label{eq:diffeotransformations}
    \begin{split}
        &\psi \mapsto \psi - A' \xi_r \ , \qquad \chi \mapsto \chi - \Phi' \xi_r \ , \qquad s \mapsto s - \tau' \xi_r  \\
        &\phi \mapsto \phi - \xi_r' - A' \xi_r \ , \qquad W \mapsto W - \xi_r - \xi' \ , \qquad E \mapsto E - \xi \ .
    \end{split}
\end{equation}
Since we have six scalars in~\eqref{eq:diffeotransformations} and two gauge functions $\xi_r$ and $\xi$, we can form four functionally independent diffeomorphism-invariant combinations. We choose\footnote{The field we denote here with $\F$ was denoted with $\widetilde{\F}$ in~\cite{Hippelainen:2026izm} while $\F$ was used for another field, but since that field does not appear in this article, we have renamed $\widetilde{\F}$ as $\F$ to simplify notation.} \begin{equation}
    \label{eq:scalarinvariants}
    \zeta = \psi - \frac{A'}{\Phi'}\chi \ , \quad \varsigma = \psi - \frac{A'}{\tau'}s \ , \quad \F = \frac{\psi}{A'} - \qty(W - E') \ , \quad \Upsilon = \phi - \pd_r \qty(\frac{\psi}{A'}) - \psi \ ,
\end{equation}
where $\zeta$ and $\varsigma$ agree with the invariants identified in~\cite{Arean:2013tja}, while $\F$ and $\Upsilon$ generalize the corresponding IHQCD combinations~\cite{Kiritsis:2006ua,Hippelainen:2026izm}. See also \cite{Berg:2005pd}.

As it turns out, not all four invariants are independent on-shell: The linearly perturbed Einstein equations imply that the combination 
\begin{equation}
    \label{eq:gconstraint}
        \psi' - A' \phi + \frac{4}{9}\Phi'\chi + \frac{x V_f \kappa \tau'}{6G}\, s = 0 \ .
\end{equation}
This can be further simplified by  using the Einstein equations in~\eqref{eq:backgroundeoms}. They imply the relation
\begin{equation}
    \label{eq:eeconstraint}
    A'' - \qty(A')^2 = - \frac{4}{9}\qty(\Phi')^2 - \frac{x}{6}\frac{V_f \kappa \qty(\tau')^2}{G} \ , 
\end{equation}
where $G$ was defined in~\eqref{eq:Gdef}. Eq.~\eqref{eq:gconstraint} therefore simplifies to \begin{equation}
        \Upsilon = - \frac{4\qty(\Phi')^2}{9\qty(A')^2}\zeta - \frac{x V_f \kappa \qty(\tau')^2}{6G\qty(A')^2}\varsigma \ ,
\end{equation}
which can be used to eliminate $\Upsilon$. 

Moreover, the remaining invariants are related by two conjugate relations for $\F$; the derivation of these equations is outlined in Appendix~\ref{subapp:conjugaterelations}. We have
\begin{equation}
\label{eq:conjugateequations}
    \begin{split}
        \F' + 3A' \F &= \frac{4\qty(\Phi')^2}{9\qty(A')^2}\zeta + \frac{x V_f \kappa \qty(\tau')^2}{6G\qty(A')^2}\varsigma  \\
        \Box \F &= - \frac{4\qty(\Phi')^2}{9\qty(A')^2}\zeta' - \frac{x V_f \kappa \qty(\tau')^2}{6G^3\qty(A')^2}\varsigma' + \mathcal{I}(r)\qty(\zeta - \varsigma) \ ,
    \end{split}
\end{equation}
where $\Box = \eta^\mn \pd_\mu \pd_\nu$, and the coefficient function is 
\begin{equation}
    \label{eq:Idef}
        \mathcal{I}(r) = -\frac{x}{6}\,\frac{V_f \kappa \Phi'\qty(\tau')^2}{G^3 \qty(A')^2}\qty[\frac{4 \Phi' \kappa \qty(\tau')^2}{9 e^{2A} A'} + \frac{1 + G^2}{2}\, \frac{\pd_\Phi \kappa}{\kappa} + G^2\, \frac{\pd_\Phi V_f}{V_f}] \ .
\end{equation}
Either of the conjugate relations determines $\F$ once $\zeta$ and $\varsigma$ are known. In practice we use the first. We write it as 
\begin{equation}
    (e^{3A} \F)' = \frac{e^{3A}}{(A')^2}\qty[\frac{4}{9}\qty(\Phi')^2 \zeta + \frac{x V_f \kappa \qty(\tau')^2}{6G} \varsigma] \ ;
\end{equation}
demanding regularity in the IR fixes the integration constant.  

For the vertex computations of Section~\ref{sec:gravitationalformfactors} we need the original fluctuation fields in terms of the gauge-invariant modes. Inverting \eqref{eq:scalarinvariants} gives
\begin{equation}
    \label{eq:invertedscalars}
    \begin{split}
        \psi &= \zeta + A' \frac{\chi}{\Phi'} \ , \qquad s = \frac{\tau'}{A'}(\zeta - \varsigma) + \tau' \frac{\chi}{\Phi'}  \\ W &= E' + \qty(\frac{\psi}{A'} - \F) \ , \qquad \phi = \frac{\psi'}{A'} + \frac{4\qty(\Phi')^2}{9(A')^2}(\psi - \zeta) + \frac{x V_f \kappa \qty(\tau')^2}{6G(A')^2}(\psi - \varsigma) \ ,
    \end{split}
\end{equation}
where the non-invariant $\chi$ and $E$ remain. By gauge invariance they must drop out of any invariant expression. 

Since the background fields depend only on the holographic coordinate, the fluctuation equations become diagonal in Fourier space, and it is sufficient to consider plane-wave modes
\begin{equation}
    \label{eq:planewavemodes}
    \begin{split}
    &\zeta(x^\mu, r) = e^{-iq \cdot x}\,\widehat{\zeta}\qty(-q^2,r) \ , \qquad \varsigma(x^\mu,r) = e^{-iq\cdot x}\,\widehat{\varsigma}\qty(-q^2,r)  \\
    &\F(x^\mu,r) = e^{-iq\cdot x}\,\widehat{\mathcal{F}}\qty(-q^2,r)\ , \qquad \hmntt(x^\mu,r) = e^{-iq\cdot x}\,\epsilon_\mn^{TT}(q)\,\widehat{h}\qty(-q^2,r) \ ,
    \end{split}
\end{equation}
where $\epsilon^{TT}_\mn(q)$ is a transverse-traceless polarization tensor, and the radial wave functions depend on the momentum only through $q^2$. 

The invariant modes $\widehat{\zeta}$ and $\widehat{\varsigma}$ are solved from a pair of coupled second-order equations first derived in~\cite{Arean:2013tja}. They are lengthy, and we reproduce them in our conventions in Appendix~\ref{subapp:conjugaterelations}, together with the decoupled equation satisfied by the tensor propagator $\widehat{h}$. Since the scalar modes are coupled, there is a two-dimensional space of IR-regular solutions, which define the bulk-to-boundary propagators. However, the only propagator we need for the computation turns out to be the propagator  which satisfies the UV boundary conditions $\widehat{\zeta}\qty(-q^2,0) = \widehat{\varsigma}\qty(-q^2,0) = 1$. The $\widehat{\F}$-component of the propagator, which we will also need when computing the form factors, is then uniquely determined from the conjugate relations by requiring IR regularity. For the tensor propagator we only need to fix the normalization, and we choose $\widehat{h}\qty(-q^2,0) = 1$.

\subsection{Linearized pion action}\label{subsec:solvingpions}

Let us then derive the linearized action for the pions. For this purpose, we write the tachyon of Section~\ref{subsec:gravityfluctuations} as 
\begin{equation}
    \label{eq:tachyonpionparametrization}
    T\qty(x^M) = \alphasymbol\qty(x^M)\, e^{i \Pi\qty(x^M)} \ , \qquad \Pi = \pi^a\qty(x^M)\, t^a \ ,
\end{equation}
where $\Omega$ is a scalar field, {\emph{i.e.}}, proportional to the unit flavor matrix, and $t^a$ are the $\liegroup{SU}{\Nf}{}$ flavor generators with the normalization $\Tr(t^a t^b) = \frac{1}{2}\delta^{ab}$, so $a,b,\ldots$ denote the adjoint flavor indices. We take the nonzero gauge fields to be $L_M = A_M$ and $R_M= -A_M$. For our purposes we can take $A_M$ to be traceless, so that we can write it as $A_\mu = A_\mu^a t^a$.

We expand the flavor action in~\eqref{eq:VQCDaction} to quadratic order in the fluctuations $(\Pi, A_M)$, keeping the metric general. This expansion also acts as the starting point of the vertex computation in Section~\ref{sec:gravitationalformfactors}. More details of this computation can be found in Appendix~\ref{subapp:linearizedpionaction}. 

We split $A_\mu = A_\mu^\perp + A_\mu^\parallel$ into a transverse part, $\pd^\mu A^\perp_\mu = 0$, and a longitudinal part, which is a gradient. The transverse part decouples from $\Pi$, and from now on $A_\mu \equiv A_\mu^\parallel$. The fluctuation action relevant for the pions then reads
\bea 
    S^{(2)}_{A\Pi} & = & -\frac{1}{2}M^3 \Nc \Tr\int \dd^5 x\, V_f\sqrt{-\det\tilde{g}}\Big[\kappa\, \alphasymbol^2\, \tilde{g}^\MN (\partial_M\Pi+2A_M) (\partial_N\Pi+2A_N)  \nonumber\\
    & & \qquad\qquad\qquad\qquad\qquad\qquad\qquad + \frac{w^2}{2}F^A_\MN \tilde{g}^{MO}\tilde{g}^{NP}F^A_{OP}\Big] \ ,\label{eq:pionaction}
\eea
where $F^A_\MN = \pd_M A_N - \pd_N A_M$, as the commutator term does not contribute at quadratic order. Note that the first term in~\eqref{eq:pionaction} causes the pseudoscalar phase $\Pi$ and the longitudinal axial field $A^\parallel_\mu$ to mix. 

We then derive the fluctuation equations for the pseudoscalars. Because the linearized action~\eqref{eq:pionaction} is already of second order in the pions, the geometry and $\Omega$ may be set to their background values. This fixes $\sqrt{-\det\tilde{g}} = e^{5A}G$ and $\alphasymbol = \tau$. It is convenient to define the following factors
\begin{equation}\label{eq:aibifactors}
    a_1 = V_f\, e^{3A} G\, \kappa \tau^2 \ , \qquad a_2 = \frac{a_1}{G^2} \ , \qquad b_1 = V_f\, e^{A} G\, w^2 \ , \qquad b_2 = \frac{b_1}{G^2} \ ,
\end{equation}
which will also appear in the fluctuation computations of Section~\ref{sec:gravitationalformfactors}. In terms of these factors, the fluctuation equations following from~\eqref{eq:pionaction} read \begin{equation}\label{eq:pioneoms}
    \begin{split}
        \pd_r\qty(a_2 \Pi') + a_1 \pd^\mu (\partial_\mu\Pi+2A_\mu) &= 0  \\
        2 a_1 (\partial_\mu\Pi+2A_\mu) - \pd_r\qty(b_2 A_\mu') &= 0  \\
        b_2 \pd^\mu A_\mu' + 2 a_2 \Pi' &= 0 \ ,
    \end{split}
\end{equation}
where these equations follow from varying with respect to $\Pi, A_\mu$, and $A_r$, respectively. We now fix a radial gauge, $A_r = 0$. To solve for the wave functions of pions, we decompose the fields into radial and four-dimensional parts, with $\Pi = \pi^a t^a$ and $A_\mu = A_\mu^a t^a$, as 
\begin{equation}
    \label{eq:pionprofiles}
    \pi^a = 2\, \psi_P(r)\, \P^a\qty(x^\mu) \ , \qquad A^{a}_\mu = - \psi_L(r)\, \pd_\mu \P^a\qty(x^\mu) \ ,
\end{equation}
where the four-dimensional part satisfies $\Box \P^a = m^2 \P^a$ with $m$ being the mass of the state.
Both fields must share the same four-dimensional mode, since the equations are coupled and the first-order constraint in~\eqref{eq:pioneoms} relates the fields. Inserting the decomposition~\eqref{eq:pionprofiles} into~\eqref{eq:pioneoms} gives the radial system~\cite{Arean:2013tja}
\begin{equation}
    \label{eq:pionradialeoms}
    \begin{split}
        \pd_r\qty(a_2 \psi_P') + m_n^2 a_1 \qty(\psi_P - \psi_L) &= 0  \\
        \pd_r\qty(b_2 \psi_L') + 4 a_1 \qty(\psi_P - \psi_L) &= 0  \\
        m_n^2 b_1 \psi_L' - 4 a_1 \psi_P' &= 0 \ .
    \end{split}
\end{equation}
Here imposing regularity in the IR and normalizability in the UV turns the system into an eigenvalue problem with a discrete tower of pseudoscalar states with masses $m=m_n$, and we identify the lowest mode with the pion, $m_1^2 = m_\pi^2$.

\section{Gravitational form factors}\label{sec:gravitationalformfactors}

We now compute the pion matrix element of the energy-momentum tensor and identify from its decomposition the gravitational form factors. By the holographic dictionary, the boundary value of the metric fluctuation sources $T^\mn$, so the matrix element follows from the part of the on-shell action that is linear in the metric fluctuation and quadratic in the pion. Therefore, in order to compute the matrix element, we vary the action~\eqref{eq:pionaction}, which is already expanded to second order in pions but is valid for a general geometry, with respect to either the tensor or all scalar modes, producing the desired third-order interaction terms. As we shall show, evaluated on the pion wave functions and the bulk-to-boundary propagators of Section~\ref{subsec:gravityfluctuations}, these cubic vertices yield the matrix element $\mel{\pi}{T^\mn}{\pi}$, from which we identify the two gravitational form factors of the pion, $\A(t)$ and $\D(t)$. Our analysis therefore expands on earlier scattering computations in similar models~\cite{Ballon-Bayona:2015wra,Ballon-Bayona:2017vlm,Amorim:2021ffr,Amorim:2021gat,Hippelainen:2026izm}.

In contrast to our earlier analysis of the proton GFFs in IHQCD~\cite{Hippelainen:2026izm}, 
where the overall normalization of the fermionic action was a free parameter fit to lattice data for the form factors, the present computation contains no such fit: all the model parameters are taken from an earlier fit to meson spectra~\cite{Amorim:2021gat}. The form factors obtained below are therefore direct predictions of the previously constructed model.

\subsection{Setting up the dictionary}\label{subsec:threepointdictionary}

We start by discussing basic definitions and details on setting up the dictionary for this specific case, where we want to compute the matrix element of the energy-momentum tensor between two pion states. 

We denote a pion with flavor $a$ and momentum $p$ by $\ket{\pi^a(p)}$, with $p^2 = -m_\pi^2$ and relativistic normalization,
\begin{equation}
    \braket{\pi^b(p')}{\pi^a(p)} = 2 E_{\vec{\boldsymbol{p}}}(2\pi)^3 \delta^3(\vec{\boldsymbol{p}}- \vec{\boldsymbol{p}}') \delta^{ab} \ .    
\end{equation}
With this normalization, the matrix element is decomposed using two gravitational form factors as (see, e.g.,~\cite{Polyakov:2018zvc}), 
\begin{equation}\label{eq:meldecomposition}
    \mel{\pi^b(p_2)}{T^\mn(0)}{\pi^a(p_1)} = \delta^{ab} \qty[2 P^\mu P^\nu \A(t) - \frac{1}{2}q^2 \Theta^\mn \D(t)] \ ,
\end{equation}
where $T^\mn(0)$ is the operator at the spacetime origin $x=0$, $\A(t)$ is related to the distribution of energy and momentum of the pion, while $\D(t)$ is related to internal stress distributions (pressure and shear). The value $\D(0)$, usually called the D-term, is defined by the zero-momentum transfer limit of the form factor. By chiral effective theory~\cite{Polyakov:1999gs}, the soft-pion theorem states that for pions, $\D(0) = -1 + O(m_\pi^2/\Lambda_\chi^2)$, where $\Lambda_\chi \approx 4\pi f_\pi$ is the chiral symmetry breaking scale~\cite{Donoghue:1991qv,Polyakov:1999gs,Hudson:2017xug}.\footnote{The usual soft-pion result assumes that no additional massless scalar contribution survives the soft limit. A massless dilaton can modify this limit; see Sections~3.2~and~3.4 of~\cite{Zwicky:2023fay}.}

The incoming and outgoing pion momenta are $p_1^\mu$ and $p_2^\mu$, from which we construct the exchanged and average momenta
\begin{equation}
    \label{eq:kinematics}
    q^\mu = p_2^\mu - p_1^\mu \ , \qquad P^\mu = \frac{1}{2}\qty(p_1 + p_2)^\mu \ .
\end{equation}
Both external legs are on-shell, $p_1^2 = p_2^2 = -m_\pi^2$, which implies $q \cdot P = 0$, and the momentum-transfer invariant is $t = -q^2 \leq 0$. The remaining scalar products are
\begin{equation}
    p_1 \cdot p_2 = - m_\pi^2 + \frac{t}{2} \ , \qquad P^2 = -m_\pi^2 + \frac{t}{4} \ .
\end{equation}
We also introduce the transverse and transverse-traceless projectors 
\begin{equation}
    \label{eq:projectors}
    \Theta^\mn(q) = \eta^\mn - \frac{q^\mu q^\nu}{q^2} \ , \qquad \TTproj = \frac{1}{2}\qty(\Theta^\alpha_\mu \Theta^\beta_\nu + \Theta^\beta_\mu \Theta^\alpha_\nu) - \frac{1}{3}\Theta^\ab \Theta_\mn \ ,
\end{equation}
which satisfy $\Theta^\mu_{\ \mu} = 3$, $q_\mu \Theta^\mn = 0$, and $q^\mu \TTproj = \eta^\mn \TTproj = 0$.

To obtain \eqref{eq:meldecomposition} holographically, we start from the QCD generating functional and follow the scattering formalism developed in~\cite{Hoyos:2019kzt,Hoyos:2020fjx,Hoyos:2022ptd}. We introduce a perturbation $\delta\eta_\mn$ of the boundary metric, $\eta_{\mu\nu}+\delta\eta_{\mu\nu}$, and a local phase\footnote{In this section, in order to simplify the notation, we do not write down explicitly the Lorentz indices for function arguments. It is understood that, for example, $f(x)$ stands for $f(x^\mu)$.} $j^a(x)$ of the quark mass matrix
\begin{equation} \label{eq:massmatrix}
    \mathsf{m}(x) = m_q e^{2 i j^a(x) t^a} \ ,
\end{equation}
where $m_q$ is the common mass of the two light quarks. The linear variation of the mass term $-\bar{q}_R \mathsf{m}q_L - \bar{q}_L \mathsf{m}^\dagger q_R$ is $j^a \O^a$ with
\begin{equation}
    \O^a = 2 m_q \bar{q}i \gamma^5 t^a q \ .
\end{equation}
The generating functional $W[\delta \eta, j]$ of connected correlators is defined by $Z[\delta\eta, j] = e^{iW[\delta\eta, j]}$, where $Z$ is the QCD partition function in the presence of the sources, which enter the QCD action as
\begin{equation}\label{eq:qcdsources}
 \delta S_\text{QCD} = \int \dd^4 x \qty[\frac{1}{2}\delta \eta_\mn T^\mn + j^a \O^a] \ .  
\end{equation}
Since metric variations are taken at fixed quark mass and gauge coupling sources, $T^\mn$ is the total QCD energy-momentum tensor. By the axial Ward identity, $\O^a$ is the divergence of the axial current $J_A^{\mu a} = \bar{q} \gamma^\mu \gamma^5 t^a q$ up to contact terms and a convention-dependent sign which drops out of the total matrix element. We set the axial current source to zero. A longitudinal axial source $a_\mu^a = \pd_\mu \omega^a$ would couple to $\pd_\mu J_A^{\mu a} \propto \O^a$. 

The corresponding gravity boundary conditions are, as $r \to 0$,
\begin{equation}
\label{eq:gravitysources}
    e^{-2A} g_\mn \to \eta_\mn + \delta \eta_\mn(x) \ , \qquad T \to C_m\, \mathsf{m}(x)\, \ell r \qty[-\log (r\Lambda_\text{UV})]^{-\rho} \ , \qquad A_\mu^a \to 0 \ ,
\end{equation}
where $T$ is the matrix-valued tachyon, $\ell$ the radius of curvature, $\Lambda_\text{UV}$ the only dimensionful parameter of the model, and $\rho$ the leading mass-running exponent. The dimensionless constant $C_m$ specifies the normalization relating a tachyon source to the QCD mass. The values of both $\rho$ and $C_m$ will be irrelevant in our work.

Since $T(x,r) = (\tau + s)e^{i \pi^a t^a}$, using~\eqref{eq:massmatrix} and~\eqref{eq:gravitysources}, we have $\pi^a(x,0) = 2 j^a(x)$, such that $C_m$ cancels as expected. We set the sources of the dilaton and of the tachyon modulus to zero, $\chi^{(0)} = s^{(0)} = 0$, since they would couple to $\Tr G_\mn G^\mn$ and shift the quark mass, and neither enters the matrix element of $T^\mn$.

The dictionary then states that 
\be\label{eq:dictionary}
 W[\delta \eta, j] = S_{\text{OS}}[\delta \eta, j]\ ,
\ee 
where $S_{\text{OS}}$ is the on-shell gravity action evaluated on the IR-regular solutions.

\subsection{Correlators on the field theory side}\label{subsec:amplitude_QCD}

We then discuss how the desired matrix element of $T^{\mu\nu}$ in~\eqref{eq:meldecomposition} is found on the field theory side of the dictionary~\eqref{eq:dictionary}. First, as usual, the coupling of $\O^a$ to a one-pion state is fixed by the pion decay constant. Since $\O^a = \pd_\mu J_A^{\mu a}$ up to contact terms, taking the divergence of $\mel{0}{J_A^{\mu a}(x)}{\pi^b(p)} = i \delta^\abe f_\pi p^\mu e^{ip\cdot x}$ and using $p^2 = -m_\pi^2$ gives
\begin{equation} \label{eq:fpidef}
    \mel{0}{\O^a(0)}{\pi^b(p)} = \delta^\abe f_\pi m_\pi^2 \ . 
\end{equation}
Notice that this construction assumes a nonzero quark mass, as appropriate for the massive pions considered here. At $m_q=0$, both $\O^a$ and its overlap with the pion vanish, so the mass phase source cannot be used to define the external pion legs. The chiral limit can instead be investigated after LSZ reduction at nonzero mass. 

The connected two- and three-point functions
\begin{equation}\label{eq:correlators}
\begin{split}
    \delta^\abe \widetilde{G}_2(p^2) &= i \int \dd^4 x e^{-ip \cdot x} \mel{0}{\mathsf{T}\qty{\O^a(x) \O^b(0)}}{0}_c \\
    \widetilde{G}_3^{\mn,a b}(p_2, p_1) &= - \int \dd^4 x \dd^4 y e^{i p_1 \cdot x - i p_2 \cdot y}\mel{0}{\mathsf{T}\qty{\O^b(y) T^\mn(0)\O^a(x)}}{0}_c \ .
\end{split}
\end{equation}
Up to contact terms these are the Fourier transforms of $\delta^2 W/\delta j \delta j$ and $2 \delta^3 W/\delta (\delta \eta) \delta j \delta j$. Here $\mathsf{T}$ denotes time-ordering. Inserting one-pion states gives
\begin{equation}
\label{eq:poles}
    \begin{split}
        \widetilde{G}_2(p^2)&=\frac{Z_\pi}{p^2+m_\pi^2-i0}+\text{regular} \\
        \widetilde G_3^{\mu\nu,ab}(p_2,p_1)&=
        \frac{Z_\pi\langle\pi^b(p_2)|T^{\mu\nu}(0)|\pi^a(p_1)\rangle}
        {(p_1^2+m_\pi^2-i0)(p_2^2+m_\pi^2-i0)}
 +\text{terms without a double pole}\ ,
    \end{split}
\end{equation}
where, using~\eqref{eq:fpidef},
\begin{equation} \label{eq:Zpires}
     Z_\pi = f_\pi^2 m_\pi^4 \ .
\end{equation}
Thus LSZ reduction means extracting the double-pole coefficient and dividing by the two external overlaps:
\begin{equation}
\label{eq:LSZ}
\langle\pi^b(p_2)|T^{\mu\nu}(0)|\pi^a(p_1)\rangle
 =\frac{1}{Z_\pi}\lim_{p_i^2\to-m_\pi^2}
 (p_1^2+m_\pi^2)(p_2^2+m_\pi^2)\widetilde G_3^{\mu\nu,ab}\ .
\end{equation}

Now, the left hand side of the dictionary in~\eqref{eq:dictionary} gives a three-point function with pion sources, and the matrix element follows only after LSZ reduction is performed. To evaluate the cubic action, the part of $W$ linear in $\delta\eta$ and quadratic in $j$ is
\begin{equation}
    \label{eq:Wcubic}
    W^{(3)}[\delta \eta, j] = \frac{1}{4}\int \dd^4 z\, \dd^4 x\, \dd^4 y\, \delta\eta_\mn(z)\, j^c(x)\, j^d(y)\, G_3^{\mn, cd}(z; x, y) \ ,
\end{equation}
where $G_3$ is the coordinate-space kernel of~\eqref{eq:correlators}, and $\frac{1}{4}$ is the product of the $\frac{1}{2}$ in~\eqref{eq:qcdsources} and the $1/2!$ of the two pion sources.
We choose the source
\begin{equation}
    \label{eq:sourcewaves}
    j^c(x) = \epsilon_1\, \delta^{ca}\, e^{ip_1\cdot x} + \epsilon_2\, \delta^{cb}\, e^{-ip_2\cdot x} \ ,
\end{equation}
at off-shell momenta, where $\epsilon_1$ and $\epsilon_2$ are bookkeeping parameters that select the incoming and outgoing pions, and $a$ and $b$ are their flavors.  Note that the structure of the flavor indices in~\eqref{eq:sourcewaves} appears to be broken, but this is just because we did not write down explicitly the dependence of the left hand side expressions on any of the source properties, {\emph{i.e.}}, the fixed momenta $p_{1,2}$ and the indices $a$ and $b$. One could assign a specific notation for such fixed indices, but following usual notation in the literature, we simply denote them by $a$ and $b$.

Let us denote the vertex~\eqref{eq:Wcubic} evaluated on the source~\eqref{eq:sourcewaves} as $W^{(3)}_\epsilon$. When we differentiate this expression with respect to $\epsilon _1$ and $\epsilon_2$, the two assignments of the plane waves to the source slots of~\eqref{eq:Wcubic} cancel the $1/2!$ factor, giving 
\begin{equation}\label{eq:Wmixed}
 \left.\frac{\partial^2W^{(3)}_\epsilon}{\partial\epsilon_1\partial\epsilon_2}\right|_{\epsilon_i=0}
 =\frac12\int\dd^4z\,\delta\eta_{\mu\nu}(z)e^{-iq\cdot z}
 \widetilde G_3^{\mu\nu,ab}(p_2,p_1)\ ,\qquad q=p_2-p_1\ .
\end{equation}
In order to obtain this expression, we used translation invariance and the fact that, due to the SU$(\Nf)$ symmetry, the three-point function is symmetric under the exchange of $a$ and $b$.
Varying~\eqref{eq:Wmixed} with respect to the metric source $\delta \eta_{\mu\nu}$ gives the final result 
\begin{equation}\label{eq:3pdictfinal}
 \widetilde G_3^{\mu\nu,ab}(p_2,p_1) =    \left.2 e^{iq\cdot z}\frac{\delta}{\delta\eta_{\mu\nu} (z)}  \frac{\partial^2W^{(3)}_\epsilon}{\partial\epsilon_1\partial\epsilon_2}\right|_{\epsilon_i=0} \ .
\end{equation}

\subsection{Correlators in gravity}\label{subsec:gravitycorr}

The analysis of the field theory correlators in Section~\ref{subsec:amplitude_QCD} pins down what we need to compute in the holographic model. Using the dictionary~\eqref{eq:dictionary}, we should evaluate the three-point function as in~\eqref{eq:3pdictfinal} and consider the limit~\eqref{eq:LSZ}. However, as seen from~\eqref{eq:poles}, fixing the normalization also requires analyzing the two-point function, which will be the topic of Section~\ref{subsec:two_point_amplitude_holo}. We then continue with the three-point function in Sections~\ref{subsec:3ptvertices}~and~\ref{subsec:amplitude_holo}.

\subsubsection{The two-point function and amputation rules}\label{subsec:two_point_amplitude_holo}

We first figure out the normalization conditions for the pion wave functions in the holographic model.
After the flavor trace, the pion action~\eqref{eq:pionaction} on the background and in the gauge $A_r = 0$ reads, using the shorthand notation defined in~\eqref{eq:aibifactors}, 
\begin{equation}\label{eq:pionactiontraced}
    S^{(2)}_{A\Pi} = -\frac{M^3 \Nc}{4}\sum_a \int \dd^4 x\, \dd r \qty[a_1\, (\partial_\mu \pi^a+2A^a_\mu) (\partial^\mu \pi^{a}+2A^{a\, \mu}) + a_2 \qty(\pd_r \pi^a)^2 + b_2\, \pd_r A^a_\mu\, \pd_r A^{a\,\mu}] \ .
\end{equation}
A normalizable pion mode~\eqref{eq:pionprofiles} solves~\eqref{eq:pionradialeoms} with $m_n = m_\pi$, is regular in the IR, and carries neither a phase nor an axial source, $\psi_P(0) = \psi_L(0) = 0$. The condition $A_\mu^a(x,0) = 0$ also fixes the residual gauge freedom of the gauge $A_r = 0$. For $m_q \neq 0$, as $r \to 0$,
\begin{equation}\label{eq:pionfalloffs}
    \psi_P,\ \psi_L \sim r^2 \ , \qquad a_1,\ a_2,\ b_2 \sim r^{-1} \ ,
\end{equation}
up to logarithms, so that boundary terms such as $a_2 \psi_P \psi_P'$ vanish at $r = 0$. We use this below to discard radial total derivatives.

The radial equations fix $\psi_{L,P}$ only up to a common normalization factor, which we keep arbitrary. Inserting the pion mode~\eqref{eq:pionprofiles} into~\eqref{eq:pionactiontraced} and separating the radial integrals gives a four-dimensional action with kinetic and mass terms,
\begin{equation}\label{eq:pion4daction}
    S_{A\Pi}^{(2)} = - \frac{1}{2}\int \dd^4 x \qty(N_\pi \pd_\mu \P^a \pd^\mu \P^a + M_\pi \P^a \P^a) \ ,
\end{equation}
with
\begin{equation}\label{eq:pionnormalization}
    N_\pi = \frac{M^3 \Nc}{2}\int \dd r \qty[4 a_1 \qty(\psi_P - \psi_L)^2 + b_2 \qty(\psi_L')^2] \ , \qquad M_\pi = 2 M^3 \Nc \int \dd r\, a_2 \qty(\psi_P')^2 \ .
\end{equation}
Using the constraint in~\eqref{eq:pionradialeoms}, $4a_2\psi_P' = m_\pi^2 b_2 \psi_L'$, and integrating the second equation by parts after multiplying it by $\psi_P - \psi_L$, we obtain
\begin{equation}\label{eq:massidentity}
    M_\pi = m_\pi^2 N_\pi \ ,
\end{equation}
where the boundary term vanishes by~\eqref{eq:pionfalloffs}. Thus, $\sqrt{N_\pi}\, \P^a$ is a canonically normalized field of mass $m_\pi$.

To relate the pole residue $Z_\pi$ to the normalization $N_\pi$ of the radial pion wave function, we compute the two-point function of $\O^a$ holographically. This requires the solution of the linearized equations~\eqref{eq:pioneoms} in the presence of a source $j^a$. In terms of its Fourier transform $\widetilde{j}^a(p) = \int \dd^4 x\, e^{-ip\cdot x} j^a(x)$, the solution is given by a pair of bulk-to-boundary propagators,
\begin{equation}\label{eq:sourcedsolution}
    \pi^a(x, r) = 2\int \frac{\dd^4 p}{(2\pi)^4}\, e^{ip\cdot x}\, K_P(p, r)\, \widetilde{j}^a(p) \ , \quad A^a_\mu(x, r) = -\int \frac{\dd^4 p}{(2\pi)^4}\, i p_\mu\, e^{ip\cdot x}\, K_L(p, r)\, \widetilde{j}^a(p) \ ,
\end{equation}
where the pair $(K_P, K_L)$ solves~\eqref{eq:pionradialeoms} with $m_n^2$ replaced by $-p^2$, is regular in the IR, and satisfies 
\be \label{eq:Knorm}
K_P(p, 0) = 1\ , \qquad K_L(p, 0) = 0 \ .
\ee 
This boundary condition is consistent with $\pi^a(x,r=0) = 2 j^a(x),\ A_\mu^a(x,r=0) = 0$. Although the pion lives in both $\Pi$ and $A_\mu^\parallel$, there is thus a single source and a single pair of propagators. 

On shell, the fluctuation equations reduce the integrand of~\eqref{eq:pionactiontraced} to a total radial derivative. The IR boundary term vanishes for a regular solution. With $A^a_\mu(x, 0) = 0$, only $\frac{M^3 \Nc}{4}\sum_a \int \dd^4 x\, a_2\, \pi^a \pd_r \pi^a$ remains at $r \to 0$. Its variations give, up to local coun\-ter\-terms,
\begin{equation}\label{eq:G2holographic}
    \begin{split}
        \expval{\O^a(x)} &= \frac{\delta W}{\delta j^a(x)} = M^3 \Nc \lim_{r \to 0} a_2\, \pd_r \pi^a(x, r)  \\
        \widetilde{G}_2(p^2) &= 2 M^3 \Nc \lim_{r\to0} a_2\, \pd_r K_P(p, r) + \text{local terms} \ .
    \end{split}
\end{equation}
Since $(K_P, K_L)$ is an IR-regular solution, it is in general obtained from any IR-regular solution by using the residual gauge symmetry $K_P \mapsto K_P + \mathrm{const}$,  $K_L \mapsto K_L + \mathrm{const}$ to remove the source of $K_L$, and by normalizing such that~\eqref{eq:Knorm} is satisfied. However, at the pion pole, the IR-regular solution is the pion mode, for which the source vanishes, so an attempt to normalize the wave function leads to a singularity. That is, $(p^2 + m_\pi^2)(K_P, K_L)$ remains finite as  $p^2 \to -m_\pi^2$.
Therefore, at the pion pole, $(p^2 + m_\pi^2)(K_P, K_L)$ solves the radial equations at $p^2 = -m_\pi^2$ with vanishing boundary values, so it is proportional to the normalizable mode,
\begin{equation}\label{eq:Kpole}
    \begin{pmatrix}
        K_P(p,r) \\
        K_L(p,r)
    \end{pmatrix} = 
    \frac{c_\pi}{p^2 + m_\pi^2}
    \begin{pmatrix}
        \psi_P(r) \\
        \psi_L(r)
    \end{pmatrix}+ \text{regular at } p^2 = -m_\pi^2 \ .
\end{equation}
There is a single coefficient $c_\pi$, since the pion is the single eigenmode of the coupled system. Comparing~\eqref{eq:G2holographic} with~\eqref{eq:poles} gives
\begin{equation}
    \label{eq:Zflux}
    Z_\pi = f_\pi^2 m_\pi^4 = 2 M^3 \Nc\, c_\pi \lim_{r\to0} a_2\, \psi_P' \ .
\end{equation}
LSZ reduction needs $c_\pi$ in terms of $N_\pi$, and the required connection follows from the radial equations~\cite{Jarvinen:2015ofa}. Applying the first and third equations of~\eqref{eq:pionradialeoms} to both $(K_P, K_L)$ and $(\psi_P, \psi_L)$, one finds
\begin{equation}\label{eq:wronskian}
    \begin{split}
        &\dv{r}\qty{4 a_2 \qty[K_P' \qty(\psi_P - \psi_L) - \psi_P' \qty(K_P - K_L)]} \\
        &= \qty(p^2 + m_\pi^2)\qty[4 a_1 \qty(K_P - K_L)\qty(\psi_P - \psi_L) + b_2\, K_L'\, \psi_L'] \ ,
    \end{split}
\end{equation}
where the $b_2$ term arises from the constraints $4a_2 K_P' = -p^2 b_2 K_L'$ and $4a_2\psi_P' = m_\pi^2 b_2 \psi_L'$. We integrate this expression radially and multiply by $M^3 \Nc/2$. The resulting IR boundary term vanishes for regular solutions, and at the UV boundary $K_P - K_L \to 1$ and $\psi_P - \psi_L \to 0$. One finds
\begin{equation}
    2 M^3 \Nc \lim_{r \to 0} a_2 \psi_P' = (p^2 + m_\pi^2) \frac{M^3 \Nc}{2}\int \dd r \qty[4 a_1 (K_P - K_L) (\psi_P - \psi_L) + b_2 K_L' \psi_L'] \ .
\end{equation}
Substituting~\eqref{eq:Kpole}, we find
\begin{equation}\label{eq:fluxnorm}
    2 M^3 \Nc \lim_{r\to0} a_2\, \psi_P' = c_\pi N_\pi \ .
\end{equation}
Together with~\eqref{eq:Zflux}, this gives $Z_\pi = c_\pi^2 N_\pi$. Choosing the sign of the profiles such that $c_\pi > 0$, we obtain
\begin{equation}\label{eq:residue}
    c_\pi = \frac{f_\pi m_\pi^2}{\sqrt{N_\pi}} \ , \qquad f_\pi = \frac{M^3 \Nc}{2\sqrt{N_\pi}}\lim_{r\to0} b_2\, \psi_L' \ .
\end{equation}

Using the relations \eqref{eq:Zpires} and \eqref{eq:residue}, we find from \eqref{eq:Kpole} that
\begin{equation}\label{eq:legrule}
    \lim_{p^2 \to -m_\pi^2}\frac{p^2 + m_\pi^2}{\sqrt{Z_\pi}} 
    \begin{pmatrix}
        K_P(p,r) \\
        K_L(p,r)
    \end{pmatrix}= \frac{1}{\sqrt{N_\pi}}
    \begin{pmatrix}
        \psi_P(r) \\
        \psi_L(r)
    \end{pmatrix}\ .
\end{equation}
Then comparing to the field theory LSZ formula~\eqref{eq:LSZ} (or equivalently to~\eqref{eq:poles}), we see that~\eqref{eq:legrule} can be understood as the amputation rule on the gravity side. For each external pion leg in the gravity generating functional, the propagator $(K_P,K_L)$ is replaced by the normalized pion wave function $(\psi_P,\psi_L)/\sqrt{N_\pi}$ when the on-shell limit is taken following~\eqref{eq:LSZ}. Note that we did not need to normalize the pion wave function $(\psi_P,\psi_L)$ explicitly; the result automatically depends on the appropriately normalized functions.

\subsubsection{Three-point vertices}\label{subsec:3ptvertices}

We are now ready to start the analysis of the three-point function in the holographic model.
By the dictionary~\eqref{eq:dictionary}, $W^{(3)}$ equals $S^{(3)}_{\rm OS}$, the part of the on-shell action linear in the metric fluctuation and quadratic in the pion, evaluated on the linear solutions sourced by $\delta\eta$ and $j$. The desired $S^{(3)}$ is obtained from the cubic terms of the fluctuated action~\eqref{eq:pionaction}. This boils down to finding the three-point vertices, where the pion field couples to the gauge-invariant modes $\zeta$, $\varsigma$, $\F$, and $\hmntt$.

Since $\hmntt$ is diffeomorphism-invariant on its own and does not mix with the scalars, its vertex follows from a short computation: fluctuating the effective metric in~\eqref{eq:pionaction} with $\tilde{g}_\mn \mapsto \tilde{g}_\mn + 2 e^{2A} \hmntt$ and keeping the longitudinal sector gives \begin{equation}\label{eq:tensorvertex}
    S^{(3)}_{\pi \pi h}
    = M^3 \Nc \Tr \int \dd^5 x\, \hmntt \qty(a_1\, (\partial^\mu \Pi+2A^\mu) (\partial^\nu\Pi+2A^\nu) + b_2\, A'^\mu A'^\nu) \ ,
\end{equation}
where the factors $a_1$ and $b_2$ were defined in~\eqref{eq:aibifactors}.

The scalar vertices require considerably more work, since all four scalars of~\eqref{eq:fluctuatedmetric}, the dilaton, and the tachyon fluctuate simultaneously, and the result must be arranged into the diffeomorphism-invariant basis. The computation, in which one fluctuates~\eqref{eq:pionaction}, substitutes the inverted relations~\eqref{eq:invertedscalars}, reduces the derivatives with the fluctuation equations~\eqref{eq:pioneoms}, and verifies that the $\chi$- and $E$-dependent terms cancel, is presented in Appendix~\ref{subapp:scalarvertexreduction}. 

The outcome is compact: the on-shell third-order action organizes into nine radial coefficient functions $c_{IJ}(r)$, which multiply the three pion bilinears, labeled by $I \in \qty{\sfU, \sfP, \sfA}$,
\begin{equation}\label{eq:bilinears}
    \sfU = \eta^\mn (\partial_\mu\Pi+2A_\mu) (\partial_\nu\Pi+2A_\nu) \ , \qquad \sfP = (\Pi')^2 \ , \qquad \sfA = \eta^\mn \qty(A_\mu^\parallel)'\qty(A_\nu^\parallel)' \ ,
\end{equation}
and the three invariants $J \in \qty{\zeta, \varsigma, \F}$. The total vertex is the sum
\begin{equation}\label{eq:scalarvertexsum}
    S_{\pi\pi s}^{(3)} = -\frac{1}{2}M^3 \Nc \Tr \int \dd^5 x \sum_{I \in \qty{\sfU, \sfP, \sfA}} \sum_{J \in \qty{\zeta,\varsigma, \F }} c_{I J} I J \ ,
\end{equation}
with $a_i$, $b_i$ as in~\eqref{eq:aibifactors} and $G$ as in~\eqref{eq:Gdef},
\begin{equation}\label{eq:scalarvertexcoefficients}
    \begin{split}
        c_{\sfU \zeta} &= \frac{\tau ' \left(\frac{2}{\tau }+\frac{\partial_\tau V_f}{V_f}\right)}{A'}a_1+\frac{4 a_2 \qty(\Phi')^2}{9(A')^2} - \qty(\frac{a_1}{A'})' +a_1 c_1+2 a_1 \\
        c_{\sfU \varsigma} &= -\frac{a_1 \tau ' \left(\frac{2}{\tau}+\frac{\partial_\tau V_f}{V_f}\right)}{A'}+\frac{x a_2 V_f \kappa \qty(\tau')^2}{6G(A')^2} + \qty(\frac{a_1 - a_2}{A'})'-a_1 c_1  \\
        c_{\sfU \F} &= a_2'-3 a_2 A'  \\
        c_{\sfP \zeta} &= \frac{\tau'}{A'}\qty(\frac{\partial_\tau V_f}{V_f} + \frac{2}{\tau})a_2 + 4a_2 - c_1 a_2 + \qty(\frac{a_2}{A'})' - 2\frac{a_2'}{A'} - \frac{4 a_2 \qty(\Phi')^2}{9G^2 (A')^2}   \\
            c_{\sfP \varsigma} &= -\frac{\tau'}{A'}\qty(\frac{\partial_\tau V_f}{V_f} + \frac{2}{\tau})a_2 + c_1 a_2 + \frac{2e^{-2A}\kappa\qty(\tau')^2 a_2'}{G^2 A'} - \qty(\frac{a_2 e^{-2A}\kappa\qty(\tau')^2}{G^2 A'})' - \frac{x a_2 V_f \kappa \qty(\tau')^2}{6G^3 (A')^2}  \\
    c_{\sfP \F} &= \frac{2a_2'}{G^2} - \qty(\frac{a_2}{G^2})' + \frac{3A' a_2}{G^2}   \\
    c_{\sfA \zeta} &= \frac{\tau'}{A'}\frac{\partial_\tau V_f}{V_f} b_2 + 2 b_2 - c_1 b_2 + \qty(\frac{b_2}{A'})' - \frac{2b_2'}{A'} - \frac{4 b_2 \qty(\Phi')^2}{9G^2(A')^2}  \\
    c_{\sfA \varsigma} &= -\frac{\tau'}{A'}\frac{\partial_\tau V_f}{V_f} b_2 + c_1 b_2 - \qty(\frac{b_2 e^{-2A}\kappa\qty(\tau')^2}{G^2 A'})' + \frac{2e^{-2A}\kappa\qty(\tau')^2 b_2'}{G^2 A'} - \frac{x b_2 V_f \kappa \qty(\tau')^2}{6G^3(A')^2}  \\
    c_{\sfA \F} &= \frac{2b_2'}{G^2} - \qty(\frac{b_2}{G^2})' + \frac{3 A' b_2}{G^2}   \\
        c_1 &= \frac{1}{G^2}\qty(\frac{x V_f \kappa \qty(\tau')^2}{6G(A')^2} + e^{-2A} \kappa \tau' \qty(\frac{\tau'}{A'})') \ .
    \end{split}
\end{equation}
The cubic action is then given by
\begin{equation} \label{eq:S3full}
    S^{(3)}_{\rm OS} =  S^{(3)}_{\pi \pi h} + S^{(3)}_{\pi \pi s}\Big|_{\rm OS} \ ,
\end{equation}
{\emph{i.e.}}, the vertices evaluated on the on-shell solutions with the sources $\delta \eta_{\mu\nu}$ and $j^a$. Note that this means that the relevant functions to insert for the pions are the propagators $(K_P,K_L)$ rather than the pion wave functions. Before inserting the on-shell wave functions of the gravitational sector explicitly, we will analyze the LSZ reduction of the pion modes.

\subsubsection{Pion amplitude}\label{subsec:amplitude_holo}

Following the field theory computation in Section~\ref{subsec:amplitude_QCD}, we first insert the sources of~\eqref{eq:sourcewaves} in $S^{(3)}_{\rm OS}$. Then the expressions for the fluctuation wave function in the action can be obtained from~\eqref{eq:sourcedsolution}. For example, the pion field becomes
\begin{equation}
    \pi^c(x,r) = 2 \epsilon_1 K_P(p_1,r)\delta^{ca}e^{ip_1\cdot x}+ 2 \epsilon_2 K_P(-p_2,r)\delta^{cb}e^{-ip_2\cdot x} \ .
\end{equation}
Let us denote the action after this replacement by $S_\epsilon^{(3)}$. Then by~\eqref{eq:3pdictfinal} and the dictionary, the holographic result for the three-point function is
\begin{equation} \label{eq:3pdictholo}
\widetilde G_3^{\mu\nu,ab}(p_2,p_1) =    \left.2 \frac{\delta}{\delta\eta_{\mu\nu} (0)}  \frac{\partial^2S^{(3)}_\epsilon}{\partial\epsilon_1\partial\epsilon_2}\right|_{\epsilon_i=0} \ .
\end{equation}

We then apply the LSZ reduction to this amplitude. The LSZ prefactor in~\eqref{eq:LSZ} factorizes over the two legs, 
\begin{equation}
    \frac{(p_1^2 + m_\pi^2)(p_2^2 + m_\pi^2)}{Z_\pi} = \prod_{i = 1,2} \frac{p_i^2 + m_\pi^2}{\sqrt{Z_\pi}} \ ,
\end{equation}
and by~\eqref{eq:legrule} each factor replaces a sourced solution $(K_P, K_L)$ by the normalizable mode $(\psi_P, \psi_L)/\sqrt{N_\pi}$. More explicitly, we can define the amputated action as
\begin{equation}\label{eq:Sampdef}
    S_{\rm amp}^{(3)} = \lim_{p_i^2\to -m_\pi^2} \frac{(p_1^2 + m_\pi^2)(p_2^2 + m_\pi^2)}{Z_\pi} S_\epsilon^{(3)} \ .
\end{equation}
Combining the above observations, we obtain the amputated action from $S_{\rm OS}^{(3)}$ in~\eqref{eq:S3full} by inserting the normalizable pion wave functions~\eqref{eq:pionprofiles} and choosing
\begin{equation}
    \label{eq:amputatedlegs}
    \P^c(x) = \frac{1}{\sqrt{N_\pi}}\qty(\epsilon_1\, \delta^{ca}\, e^{ip_1\cdot x} + \epsilon_2\, \delta^{cb}\, e^{-ip_2\cdot x}) 
\end{equation}
for the spacetime-dependent wave functions.
Note that the metric fluctuation is not amputated and remains the solution sourced by $\delta\eta_\mn$. 

In summary, the final vertex rule is
\begin{equation}\label{eq:vertexrule}
    \mel{\pi^b(p_2)}{T^\mn(0)}{\pi^a(p_1)} = 2\, \frac{\delta}{\delta(\delta\eta_\mn(0))} \frac{\pd^2 S^{(3)}_{\rm amp}}{\pd \epsilon_1 \pd \epsilon_2} \ ,
\end{equation}
where $S^{(3)}_{\rm amp}$ is evaluated with the pion fields~\eqref{eq:pionprofiles}, with $\P^c$ given by~\eqref{eq:amputatedlegs}, and with the metric fluctuations sourced by $\delta\eta_\mn$. Local counterterms have no double pole and do not contribute.

\subsection{Extracting the form factors}\label{subsec:three_point}

A few final details need to be clarified before we can compute the GFFs.
In the cubic action~\eqref{eq:Sampdef} the pion enters through flavor traces of bilinears such as $\Tr(\pd^\mu \P\, \pd^\nu \P)$, with $\P = \P^c t^c$. Inserting~\eqref{eq:amputatedlegs} and keeping the coefficient of $\epsilon_1\epsilon_2$, denoted by $(\cdot)_{\epsilon_1\epsilon_2}$, gives
\begin{equation}\label{eq:momentumrules}
    \begin{split}
        \qty(\Tr \pd^\mu \P\, \pd^\nu \P)_{\epsilon_1\epsilon_2} &= \frac{\delta^{ab}}{N_\pi}\qty(P^\mu P^\nu - \frac{q^\mu q^\nu}{4})e^{-iq\cdot x}  \\
        \qty(\Tr \pd \P \cdot \pd \P)_{\epsilon_1\epsilon_2} &= \frac{\delta^{ab}}{N_\pi}\, p_1 \cdot p_2\, e^{-iq \cdot x} \ , \qquad \qty(\Tr \P^2)_{\epsilon_1\epsilon_2} = \frac{\delta^{ab}}{N_\pi}\, e^{-iq\cdot x} \ ,
    \end{split}
\end{equation}
which include both assignments of the external legs and $\Tr(t^a t^b) = \frac{1}{2}\delta^{ab}$. 

To evaluate the vertex, we express the boundary values of the gravity fluctuations through $\delta \eta_\mn$. At the boundary, the metric fluctuation term in the square brackets of~\eqref{eq:fluctuatedmetric} reduces to 
\begin{equation}\label{eq:boundarymetric}
    2 \psi^{(0)} \eta_\mn + 2\, \pd_\mu \pd_\nu E^{(0)} + 2\, h_\mn^{TT,(0)} = \delta \eta_\mn \ ,
\end{equation}
where the superscript $(0)$ denotes a boundary value. Contracting with $\eta^\mn$, $q^\mu q^\nu/q^2$, and $\TTproj$ in momentum space and inverting gives, with the projectors~\eqref{eq:projectors} and tildes denoting Fourier transforms, $\widetilde{f}(q) = \int \dd^4 x\, e^{-iq\cdot x} f(x)$,
\begin{equation}\label{eq:boundaryvalues}
    \widetilde{\psi}^{(0)}(q) = \frac{1}{6}\Theta^\mn(q)\, \widetilde{\delta \eta}_\mn(q) \ , \qquad \widetilde{h}_\mn^{TT,(0)}(q) = \frac{1}{2}\TTproj\, \widetilde{\delta \eta}_\ab(q) \ .
\end{equation}
The boundary value $E^{(0)}$ is obtained similarly, but it does not contribute to the final result. 

Since the sources $\chi^{(0)}$ and $s^{(0)}$ vanish, the boundary values of the invariants collapse by~\eqref{eq:scalarinvariants} onto the single scalar source\footnote{Strictly speaking, $\chi$ and $s$ in~\eqref{eq:scalarinvariants} retain their VEV contributions, but by UV power counting their contributions to the invariants, $(A'/\Phi')\chi$ and $(A'/\tau')s$, vanish at the boundary when $m_q \neq 0$; see Appendix~\ref{subapp:boundaryasymptotics}.}
\begin{equation}\label{eq:invariantboundaryvalues}
    \widetilde \zeta^{(0)} = \widetilde \varsigma^{(0)} = \widetilde \psi^{(0)} \ ,
\end{equation}
while the boundary behavior of $\F$ follows from the conjugate relations~\eqref{eq:conjugateequations}. Therefore, the full sourced solutions for the invariant modes of~\eqref{eq:planewavemodes} are given by 
\begin{align} \label{eq:finalpropagators}
    \zeta(x,r) &= \int \frac{\dd^4q}{(2\pi)^4}\,  \widetilde{\psi}^{(0)}(q)\,  \widehat{\zeta}\qty(-q^2,r)\,e^{iq\cdot x} \nonumber\\
    \varsigma(x,r) &= \int \frac{\dd^4q}{(2\pi)^4}\,  \widetilde{\psi}^{(0)}(q)\,  \widehat{\varsigma}\qty(-q^2,r)\,e^{iq\cdot x} \nonumber\\
    \F(x,r) &= \int \frac{\dd^4q}{(2\pi)^4}\,  \widetilde{\psi}^{(0)}(q)\,  \widehat{\F}\qty(-q^2,r)\,e^{iq\cdot x} \nonumber\\
    \hmntt(x,r) &= \int \frac{\dd^4q}{(2\pi)^4}\, \widetilde{h}_\mn^{TT,(0)}(q)\,  \widehat{h}\qty(-q^2,r)\,e^{iq\cdot x} \ ,
\end{align}
where $\widehat{\zeta}$, $\widehat{\varsigma}$, and  $\widehat{\F}$ are the components of the scalar bulk-to-boundary propagator satisfying the boundary condition $\widehat{\zeta}\qty(-q^2,0)=1=\widehat{\varsigma}\qty(-q^2,0)$, as discussed in Section~\ref{subsec:gravityfluctuations}, while $\widehat{h}$ is the tensor propagator similarly normalized to one  at the boundary.

The natural split of the vertex on the gravity side is not the one in~\eqref{eq:meldecomposition}, but a split into transverse-traceless (TT) and transverse-traceful parts. We therefore extract the trace part of $P^\mu P^\nu$, 
\begin{equation}
    P^\mu P^\nu = \qty(P^\mu P^\nu - \frac{P^2}{3}\Theta^\mn) + \frac{P^2}{3}\Theta^\mn \ ,
\end{equation}
and reorganize the matrix element as 
\begin{equation}\label{eq:melTTsplit}
    \begin{split}
        \mel{\pi^b(p_2)}{T^\mn(0)}{\pi^a(p_1)} &= \delta^{ab} \qty[2 \A(t)\, \qty(P^\mu P^\nu - \frac{P^2}{3}\Theta^\mn) + \S(t)\, \Theta^\mn]  \\
        \S(t) &= \frac{2P^2}{3}\A(t) - \frac{q^2}{2}\D(t) \ .
    \end{split}
\end{equation}
Hence, the tensor vertex~\eqref{eq:tensorvertex} determines $\A(t)$ alone, while the scalar vertex~\eqref{eq:scalarvertexsum} determines $\S(t)$, which contains contributions from both form factors. Once both vertices are known, $\D(t)$ is obtained by inverting the relation in~\eqref{eq:melTTsplit}.

We are now ready to evaluate the vertices, starting with the tensor vertex.  Inserting the external legs, the propagator in~\eqref{eq:finalpropagators} with the boundary decomposition~\eqref{eq:boundaryvalues}, and the momentum rules~\eqref{eq:momentumrules} into~\eqref{eq:vertexrule}, and comparing with the TT part of~\eqref{eq:melTTsplit}, we can identify the first form factor: 
\begin{equation}\label{eq:Atidentification}
    \A(t) = \frac{M^3 \Nc}{2 N_\pi} \int \dd r \qty[4 a_1 \qty(\psi_P - \psi_L)^2 + b_2 \qty(\psi_L')^2]\, \widehat{h}(t,r) \ .
\end{equation}
At zero momentum transfer the tensor propagator is exactly unity, $\widehat{h}(0,r) = 1$, and the integral in~\eqref{eq:Atidentification} reduces to the normalization integral in~\eqref{eq:pionnormalization}. Therefore, $\A(0) = \frac{N_\pi}{N_\pi} = 1$ for any normalization of the profiles. That is, as a useful cross-check, the momentum sum rule is satisfied automatically. 

Since all three scalar invariants carry the same boundary factor $\widetilde{\psi}^{(0)}(q)$ in~\eqref{eq:finalpropagators}, their propagators enter the vertex~\eqref{eq:scalarvertexsum} only through the combinations 
\begin{equation}\label{eq:scriptCcombinations}
    \C_I(t,r) = c_{I\zeta}(r)\, \widehat{\zeta}(t,r) + c_{I\varsigma}(r)\, \widehat{\varsigma}(t,r) + c_{I\F}(r)\, \widehat{\mathcal{F}}(t,r) \ , \qquad I \in \qty{\sfU, \sfP, \sfA} \ .
\end{equation}
Inserting \eqref{eq:amputatedlegs}, \eqref{eq:momentumrules}, and \eqref{eq:finalpropagators} into the scalar vertex~\eqref{eq:scalarvertexsum} gives
\begin{equation}
  \frac{\pd^2 S^{(3)}_{\pi\pi s}}{\pd\epsilon_1\pd\epsilon_2} = -\frac{M^3\Nc}{4N_\pi} \delta^{ab} \widetilde\psi^{(0)} \Sigma\qty(t) \ ,
\end{equation}
with the scalar amplitude 
\begin{equation}\label{eq:scalaramplitude}
    \Sigma(t) = \int \dd r\, \qty{2\, p_1 \cdot p_2 \qty[4 \qty(\psi_P - \psi_L)^2\, \C_\sfU(t,r) + \qty(\psi_L')^2\, \C_\sfA(t,r)] + 8 \qty(\psi_P')^2\, \C_\sfP(t,r)} \ ,
\end{equation}
where $p_1 \cdot p_2 = -m_\pi^2 + t/2$. Since $\widetilde\psi^{(0)} = \frac{1}{6} \Theta^\mn \widetilde{\delta\eta}_\mn$ by~\eqref{eq:boundaryvalues}, the variation in~\eqref{eq:vertexrule} gives a term proportional to $\Theta^\mn$. We then compare with the $\Theta^\mn$-term of the decomposition~\eqref{eq:melTTsplit} and find
\begin{equation}\label{eq:Stidentification}
    \S(t) = - \frac{M^3 \Nc}{12 N_\pi}\, \Sigma(t) \ .
\end{equation}
Finally, inverting~\eqref{eq:melTTsplit} for $\D(t)$ gives
\begin{equation}\label{eq:Dtidentification}
    \D(t) = - \frac{1}{3}\A(t) + \frac{4 m_\pi^2}{3 t}\A(t) - \frac{M^3 \Nc}{6 N_\pi t}\, \Sigma(t) \ .
\end{equation}
The last two terms of~\eqref{eq:Dtidentification} carry explicit poles at $t = 0$, so regularity of $\D(t)$ requires exact cancelation between them, which amounts to 
\begin{equation}\label{eq:Sigma0}
    \Sigma(0) = \frac{8 m_\pi^2 N_\pi}{M^3 \Nc} \ .
\end{equation}
At $t = 0$ the coupled system~\eqref{eq:coupledscalarequations} is solved by $\widehat \zeta(0,r) = \widehat \varsigma(0,r) = 1$, while $\widetilde \F(0,r)$ does not vanish; it is fixed by the first conjugate relation~\eqref{eq:conjugateequations} at $t=0$. Inserting this into \eqref{eq:scalaramplitude} and massaging the result gives~\eqref{eq:Sigma0} analytically; see Appendix~\ref{subapp:regularityofD} for details. Therefore, $\D(t)$ is regular at vanishing momentum transfer.

\subsection{Results for the form factors}\label{subsec:formfactorcomputation}

\begin{figure}[!htb]
    \centering
    \includegraphics[width=0.8\textwidth]{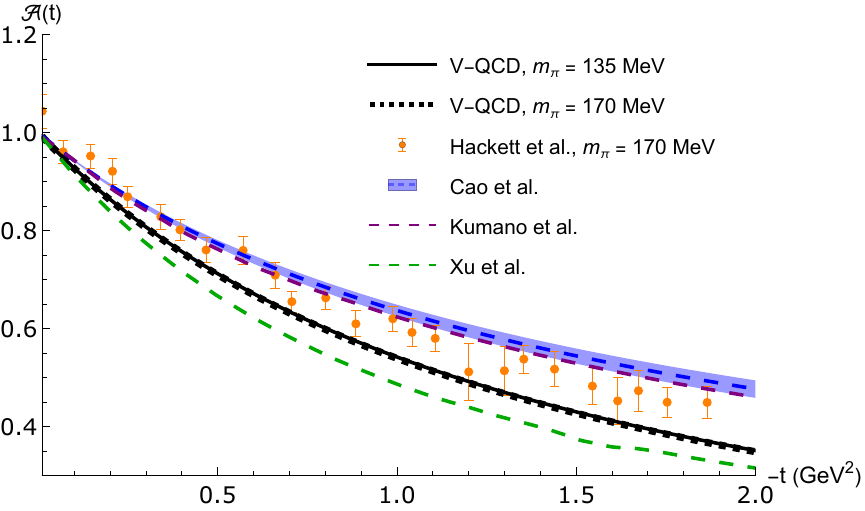}
    \caption{The pion form factor $\A(t)$. The black solid and dotted curves are our predictions at $m_\pi \approx 135$~MeV and $m_\pi \approx 170$~MeV, respectively. The orange points are the lattice data of~\cite{Hackett:2023nkr} at $m_\pi \approx 170$~MeV. The blue dashed curve and the shaded band show the central value and uncertainty of the dispersive analysis of~\cite{Cao:2024zlf}. The purple dashed curve shows the normalized quark-sector form factor from the Belle/GDA analysis~\cite{Kumano:2017lhr}, and the dark green dashed curve shows the result obtained using Dyson--Schwinger methods~\cite{Xu:2023izo}.}
    \label{fig:Atplot}
\end{figure}

\begin{figure}[!htb]
    \centering
    \includegraphics[width=0.8\textwidth]{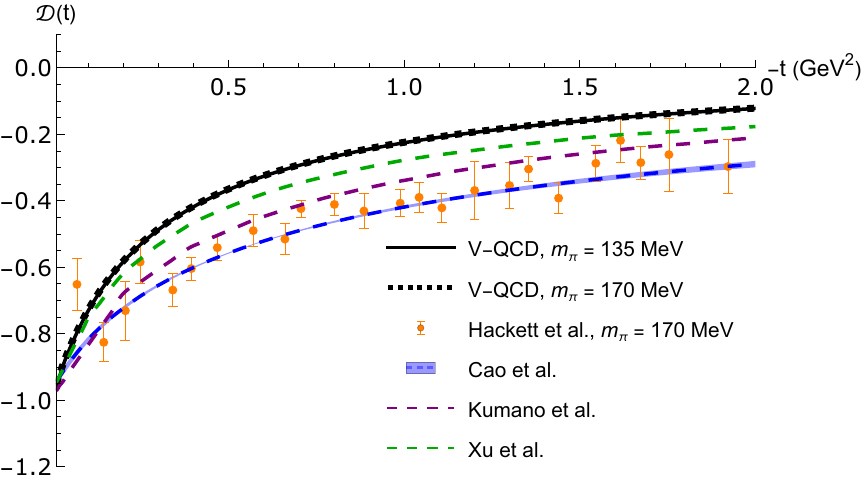}
    \caption{Results for the form factor $\D(t)$ of the pion, with curves and data as in Fig.~\ref{fig:Atplot}. The Kumano et al. curve is the normalized quark-sector $\D$ form factor shape, rescaled to $\D(0)=-1$ for comparison of momentum dependence.}
    \label{fig:Dtplot}
\end{figure}

At this point, every ingredient entering the formulas~\eqref{eq:Atidentification} and~\eqref{eq:Dtidentification} is fixed, and the form factors can be evaluated using the potentials and backgrounds specified in Appendix~\ref{app:potentials} based on the fit to QCD spectrum data in~\cite{Amorim:2021gat}. The form factors are shown in Figs.~\ref{fig:Atplot} and~\ref{fig:Dtplot} in the range $0 \leq -t \leq 2$~GeV$^2$ both at the physical pion mass $m_\pi \approx 135$~MeV and at $m_\pi \approx 170$~MeV for a direct comparison with lattice data.

Although the poles in~\eqref{eq:Dtidentification} cancel exactly, numerically $\D(t)$ is the small difference of two terms that each grow as $1/t$, so its relative error grows as $t \to 0$. We therefore fit a quadratic polynomial to the values at the four smallest nonzero momentum transfers of our grid, which gives
\begin{equation}\label{eq:Dtermvalue}
    \D(0) \approx -0.984 \ .
\end{equation}
We have also computed the D-term when the pion mass is 170~MeV and found it to be $\D(0) \approx -0.978$.

Let us then compare our results (black solid and dashed curves in Figs.~\ref{fig:Atplot} and~\ref{fig:Dtplot}) with the other results. The pion has received considerably less attention than the proton, and determinations of both of its gravitational form factors with nonzero quark masses remain scarce. The lattice study of Hackett et al.~\cite{Hackett:2023nkr} provides both form factors, including the decomposition of form factors into quark and gluon contributions; the data points in Figs.~\ref{fig:Atplot}~and~\ref{fig:Dtplot} correspond to their total form factor data. The total pion $\D$ form factor data of~\cite{Hackett:2023nkr} have also been analyzed in terms of a $\sigma/f_0(500)$ pole supplemented by a background contribution~\cite{Stegeman:2025sca}. For their central fit at $m_\pi\approx 170$ MeV, the parameters of~\cite{Stegeman:2025sca} imply $\D(0)=-1+\hat b=-0.902(81)$, consistent with both lattice data and the soft-pion expectation within uncertainties. The dispersive analysis results of Cao et al.~\cite{Cao:2024zlf}, given as the blue dashed curves, come with particularly small error estimates, shown as the light blue bands. On the experimental side, Kumano et al.~\cite{Kumano:2017lhr} extract quark-sector form factors from Belle Collaboration data on $\gamma^* \gamma \to \pi^0 \pi^0$ using generalized distribution amplitudes. Since the separation into quark and gluon contributions is renormalization-scale and scheme dependent, while our calculation concerns the total form factors, we use these results only for a qualitative comparison. In addition, because their analysis does not fix the overall normalization of the D-term, the plotted curve is normalized with $\D(0)=-1$ rather than representing an experimental determination of the D-term magnitude. We also show, in Figs.~\ref{fig:Atplot} and~\ref{fig:Dtplot}, the predictions  of Xu et al.~\cite{Xu:2023izo} obtained with Dyson--Schwinger methods. Our results for the $t$-dependence of the form factors are in qualitative agreement with other methods, including the precise dispersive results and lattice analysis. However, the decrease in the magnitudes of the form factors with increasing $-t$ is somewhat faster than what the dispersive and lattice results indicate, in particular for $\D(t)$. Finally, one might speculate that the tension between the dispersive and lattice results for $\A(t)$ in Fig.~\ref{fig:Atplot} is due to the unphysical pion mass of the lattice analysis. Our results partially support this: they change in the correct direction as the pion mass is varied, but the magnitude of the change is much smaller than the difference between the dispersive curve and lattice data.

\section{Mechanical properties}\label{sec:mechanicalproperties}

The gravitational form factors encode the matrix elements of the energy-momentum tensor and can be used to construct spatial distributions associated with the energy and internal forces of hadrons~\cite{Polyakov:2002yz}. In particular, the form factor $\D(t)$ determines the distributions conventionally called pressure and shear, while both $\A(t)$ and $\D(t)$ enter the energy density. Their interpretation as local mechanical properties of the pion is, however, not unambiguous. Localizing such a light particle on the scale of its size involves substantial relativistic effects, which complicate the identification of these distributions with classical local pressure and energy density~\cite{Jaffe:2020ebz,Epelbaum:2022fjc}.

In V-QCD, as in QCD without electroweak interactions, the pion is stable. In nature its lifetime, at least $\sim 10^{-17}$~s for the $\pi^0$, exceeds the strong-interaction time scale $\sim 10^{-23}$~s by many orders of magnitude, so the pion can be treated as a stationary state while its structure is probed.

For comparison, we compute the pressure, shear, and energy density distributions also from the monopole fits of~\cite{Hackett:2023nkr} to their lattice form factors. In Section~\ref{subsec:pressureshearandenergydensity} we compute the pressure, shear, and energy density distributions, and in Section~\ref{subsec:pionradii} the associated radii. As we saw, the holographic predictions for $\A(t)$ and $\D(t)$ were quite similar at $m_\pi \approx 135$~MeV and $m_\pi \approx 170$~MeV. However, the mechanical distributions also depend explicitly on $m_\pi$ through the relativistic normalization factor. The $\rho^2$-weighted densities differ by at most a few percent between these two masses, while radii are more sensitive. Hence, we compute everything in this section only for the form factors at the physical pion mass, except the radii, which we compute for both holographic profiles.

\subsection{Pressure, shear, and energy density}\label{subsec:pressureshearandenergydensity}

Let us set up the relevant definitions, following the discussion of spin-0 hadrons in~\cite{Polyakov:2018zvc}. We work in the Breit frame, where $q^\mu \equiv p_2^\mu-p_1^\mu= (0, \va{q})$, with $p_i$ being the momenta of the incoming or outgoing pions. Therefore, throughout this section, $q$ denotes $|\va{q}|$, and the form factors are written as functions of $-q^2$. 
Since in the Breit frame $P^i = 0$, the spatial components of the matrix element~\eqref{eq:meldecomposition} depend on $\D$ alone,
\begin{equation}\label{eq:spatialEMT}
    \mel{\pi^b(p_2)}{T^\ij(0)}{\pi^a(p_1)} = \frac{\delta^{ab}}{2}\qty(q^i q^j - \delta^\ij q^2) \D(-q^2) \ .
\end{equation}

We then define the static distribution as the Fourier transform of this matrix element without the flavor factor,
\begin{equation}\label{eq:Tijdef}
    T_\ij(\va{\rho}) = \int \frac{\dd^3 q}{(2\pi)^3\, 2E(q)}\, e^{-i\va{q}\cdot\va{\rho}}\, \frac{1}{2}\qty(q_i q_j - \delta_\ij q^2)\D(-q^2) = \frac{1}{2}\qty(\delta_\ij \nabla^2 - \pd_i \pd_j)\widetilde{\D}(\rho) \ ,
\end{equation}
where $\va{\rho}$ is the position vector measured from the center of the pion, $\rho$ denotes its length, and
\begin{equation}\label{eq:Dtilderelativistic}
    \widetilde{\D}(\rho) = \int \frac{\dd^3 q}{(2\pi)^3\, 2 E(q)}\, e^{-i \va{q} \cdot \va{\rho}}\, \D(-q^2) \ , \qquad E(q) = \sqrt{m_\pi^2 + \frac{q^2}{4}} \ .
\end{equation}
The transformed function $\widetilde{D}$ depends on $\va{\rho}$ only through $\rho$. The factor $1/(2E(q))$ in (\ref{eq:Tijdef}) follows from the relativistic normalization of states in the Breit-frame prescription for spin-zero hadrons~\cite{Polyakov:2018zvc,Hudson:2017xug}. Our treatment of the large-momentum tails in the Fourier transforms is described in Appendix~\ref{app:tailfitsandsystematics}.

The spatial components of the static energy-momentum tensor can be decomposed into pressure and shear forces $p(\rho)$ and $s(\rho)$,
\begin{equation}\label{eq:Tijdecomposition}
    T_\ij(\va{\rho}) = p(\rho)\, \delta_\ij + s(\rho) \qty(\frac{\rho_i \rho_j}{\rho^2} - \frac{1}{3}\delta_\ij) \ .
\end{equation}
In terms of $\widetilde{\D}$, the pressure and shear profiles read
\begin{equation}\label{eq:pressuresheardefs}
    p(\rho) = \frac{1}{3 \rho^2} \dv{}{\rho} \qty(\rho^2 \dv{\widetilde{\D}}{\rho}) \ , \qquad  s(\rho) = -\frac{1}{2}\rho \dv{}{\rho} \qty(\frac{1}{\rho} \dv{\widetilde{\D}}{\rho}) \ .
\end{equation}
By carrying out the angular integrals, we can represent the mechanical properties directly in terms of $\D(t)$:
\begin{equation}\label{eq:pressureshearintegrals}
    p(\rho) = -\frac{1}{12 \pi^2 \rho} \int_0^\infty \dd q\, q^3\, \frac{\sin(q\rho)}{E(q)}\, \D(-q^2) \ , \qquad s(\rho) = -\frac{1}{8 \pi^2} \int_0^\infty \dd q\, q^4\, \frac{j_2(q\rho)}{E(q)}\, \D(-q^2) \ ,
\end{equation}
where $j_2$ is the spherical Bessel function of order two. We evaluate these integrals numerically and check that the resulting profiles agree with the ones computed from $\widetilde{D}$.

The normal and tangential pressures are defined by projecting the spatial part of the energy-momentum tensor with radial and tangential unit vectors. These pressures can be defined directly in terms of the total pressure and shear profiles, either with the Fourier transform or in terms of an integral weighted by a spherical Bessel function,
\begin{equation}\label{eq:normaltangential}
    p_\rho(\rho) = p(\rho) + \frac{2}{3} s(\rho) = -\frac{1}{4\pi^2 \rho}\int_0^\infty \dd q\, q^3\, \frac{j_1(q \rho)}{E(q)}\, \D(-q^2) \ , \qquad p_\theta(\rho) = p(\rho) - \frac{1}{3}s(\rho) \ .
\end{equation}
Finally, the energy density $\epsilon(\rho)$ is the $T^{00}$ component of the static distribution.
Including the relativistic normalization as above, it reads
\begin{equation}\label{eq:energydensity}
    \epsilon(\rho) = \frac{m^2_\pi}{2\pi^2 \rho} \int_0^\infty \dd q\, \frac{q \sin(q\rho)}{E(q)} \qty[ \A(-q^2) + \frac{q^2}{4 m_\pi^2}\qty(\A(-q^2) + \D(-q^2))] \ .
\end{equation}
Replacing the momentum-dependent normalization factor $1/(2E(q)))$ by its value at zero momentum, $1/(2m_\pi)$, gives the static approximation. The expression entering the Fourier transform then becomes
\be
 m_\pi \qty[\A(t) - \frac{t}{4 m_\pi^2}\qty(\A(t) + \D(t))] \ , \ \qquad t=-q^2 \ . 
\ee

Evaluating these Fourier transforms requires the form factors at all momentum transfers, but since it is hard to extend the holographic computation to asymptotically large momentum transfers, the numerical profiles are eventually continued by power-law tails with the perturbative QCD falloff $\A(t), \D(t) \sim 1/t$~\cite{Tong:2021ctu}. We detail their construction in Appendix~\ref{app:tailfitsandsystematics}.

One consequence of the $1/t$ asymptotics is worth stating explicitly. For the proton (see, {\emph{e.g.}},~\cite{Hippelainen:2026izm}), the faster $1/t^3$ falloff~\cite{Brodsky:1973kr, Tanaka:2018wea} of $\D(t)$ makes the combinations $\rho^2 p$, $\rho^2 s$, and $\rho^2 \epsilon$ vanish at the origin but for the pion they approach finite nonzero constants, set by the coefficients of the leading tails, $\A(t) \simeq \frac{\A_1}{-t}$ and $\D(t) \simeq \frac{\D_1}{-t}$ as $-t \to \infty$, determined in Appendix~\ref{app:tailfitsandsystematics}:
\begin{equation}\label{eq:centralvalues}
    \rho^2 p \to -\frac{\D_1}{6\pi^2} \ , \qquad \rho^2 s \to -\frac{\D_1}{2\pi^2} \ , \qquad \rho^2 \epsilon \to \frac{\A_1 + \D_1}{4\pi^2} \ , \qquad \text{as } \rho \to 0 \ .
\end{equation}

\begin{figure}[!htb]
    \centering
    \includegraphics[width=0.8\textwidth]{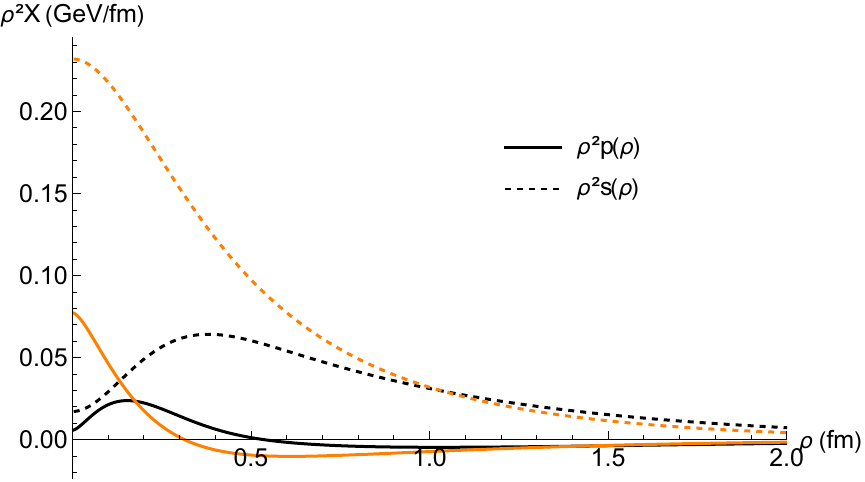}
    \caption{The $\rho^2$-weighted pressure (solid) and shear force (dashed) distributions of the pion computed from our holographic result (black) and the monopole fits of~\cite{Hackett:2023nkr} (orange) to lattice data at $m_\pi \approx 170$~MeV.}
    \label{fig:pressureshear}
\end{figure}

\begin{figure}[!htb]
    \centering
    \includegraphics[width=0.8\textwidth]{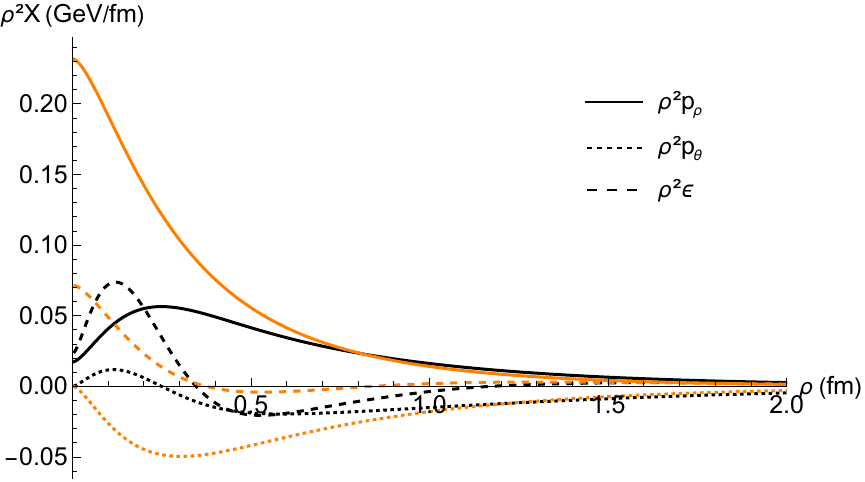}
    \caption{The $\rho^2$-weighted distributions of the normal pressure $p_\rho$, tangential pressure $p_\theta$, and energy density $\epsilon$ of the pion, computed from our holographic result (black) and the monopole fit of~\cite{Hackett:2023nkr} (orange).}
    \label{fig:normaltangentialenergy}
\end{figure}

The pressure and shear distributions~\eqref{eq:pressureshearintegrals} are shown in Fig.~\ref{fig:pressureshear}, and the normal pressure, tangential pressure, and energy density distributions in Fig.~\ref{fig:normaltangentialenergy}. For comparison, we also show the distributions computed from the monopole fits of~\cite{Hackett:2023nkr} to their lattice form factors at $m_\pi \approx 170$~MeV,
\begin{equation}
\begin{split}
    \A_\text{mon}(t) &= \frac{0.546}{1 - \frac{t}{(1.129 \text{ GeV})^2}} + \frac{0.481}{1 - \frac{t}{(1.262 \text{ GeV})^2}}  \\
    \D_\text{mon}(t) &= - \frac{0.596}{1 - \frac{t}{(0.677 \text{ GeV})^2}} - \frac{0.304}{1 - \frac{t}{(1.44 \text{ GeV})^2}} \ .
\end{split}
\end{equation}
As for our profiles, the integrals are evaluated numerically up to $-t = 6.4$~GeV$^2$. Beyond this point, we use the large-$|t|$ expansion of the monopoles in powers of $1/t$ and integrate analytically.

Our Breit-frame pressure in Fig.~\ref{fig:pressureshear} is positive at small $\rho$, crosses zero at $\rho \approx 0.54$~fm, and becomes negative at larger $\rho$. The pressure obtained from the lattice fits is otherwise similar to our result, but in the small $\rho$ regime, unlike the holographic result, it is monotonic. The pressure given by the lattice fit changes sign at a smaller radius than the holographic prediction, at $\rho \approx 0.32$~fm. 

The $\rho^2$-weighted shear distribution is positive for both the holographic and lattice profiles. It peaks at $\rho \approx 0.39$~fm in the holographic case, while the lattice result decreases monotonically and is largest at $\rho = 0$.

In Fig.~\ref{fig:normaltangentialenergy}, our normal pressure is positive for all $\rho$, while the tangential pressure changes sign already at $\rho \approx 0.25$~fm. For the lattice fits, the tangential pressure is negative across the whole range of $\rho$. 

The $\rho^2$-weighted energy density, also shown in Fig.~\ref{fig:normaltangentialenergy}, is  strongly peaked at small $\rho$ for our holographic profile, but in contrast to the proton case \cite{Hippelainen:2026izm}, it is not positive definite: $\rho^2 \epsilon$ is negative in the intermediate region $\rho \approx 0.35$--$1.17$~fm  before going to zero from above at large $\rho$.\footnote{As a consistency check, the energy density integrates to the pion mass, $\int \dd^3 \rho\, \epsilon(\rho) = m_\pi$, which our profiles satisfy.}  Interpreting the Breit-frame Fourier transforms as static spatial densities requires that the position of the hadron can be fixed to a precision much better than its size without generating relativistic effects, which holds only when its Compton wavelength is much smaller than its size~\cite{Jaffe:2020ebz}; light-front densities~\cite{Freese:2021czn} and a definition of spatial densities free of this restriction~\cite{Epelbaum:2022fjc} have also been proposed as alternatives. For the pion, $1/m_\pi \approx 1.46$~fm exceeds the radii obtained in Section~\ref{subsec:pionradii}, so the interpretation of these Breit-frame profiles as classical static densities is not justified~\cite{Polyakov:2018zvc,Jaffe:2020ebz}. The energy density obtained from the lattice fits also has a negative region, at $0.37 \lesssim \rho \lesssim 0.84$~fm, so the feature is not specific to our model. The negative region should therefore not be interpreted directly as a region of negative classical mass density in a localized pion. The integral of $\epsilon(\rho)$ continues to satisfy the pion mass sum rule.

\subsection{Pion radii}\label{subsec:pionradii}

With the distributions at hand, we may finally estimate the size of the pion. A radius is defined as the square root of the second moment of a given density,
\begin{equation}\label{eq:radiusdefinition}
    \rho_X = \sqrt{\lara{\rho^2}_X} \ , \qquad \lara{\rho^2}_X = \frac{\int \dd^3 \rho\, \rho^2\, X(\rho)}{\int \dd^3 \rho\, X(\rho)} \ ,
\end{equation}
where for the mass radius the weight $X$ is the energy density~\eqref{eq:energydensity}, and for the mechanical radius the normal pressure $p_\rho$ of~\eqref{eq:normaltangential}, which is the natural weight for a radial moment~\cite{Polyakov:2018zvc}. The mechanical radius can also be computed directly from $\D(t)$,
\begin{equation}\label{eq:mechradius}
    \lara{\rho^2}_\text{mech} = \frac{3 \D(0)}{m_\pi \int_{0}^\infty \dd q \ q  \D(-q^2)/E(q)} \ .
\end{equation}

The estimates of the radii are
\begin{equation}\label{eq:radiiresults}
    \rho_\text{mass} \approx 1.32 \tx{fm} \ , \qquad \rho_\text{mech} \approx 0.86 \tx{fm} \ ,
\end{equation}
so that, as for the proton \cite{Hippelainen:2026izm}, the mass radius exceeds the mechanical one, with the energy distribution extending well beyond the region where the internal forces are concentrated. The radii are more sensitive to the pion mass than the distributions: with $m_\pi = 170$~MeV, the holographic predictions are $\rho_\text{mass} \approx 1.07$~fm and $\rho_\text{mech} \approx 0.79$~fm. We compute the radii estimates also for the lattice parametrization of~\cite{Hackett:2023nkr} and find $\rho_\text{mass} \approx 0.99$~fm and $\rho_\text{mech} \approx 0.53$~fm. At the same pion mass, the lattice mass radius is thus close to ours, while the lattice mechanical radius is about a third smaller.

\section{Conclusions and outlook}\label{sec:conclusions}

In this article we computed the total gravitational form factors of the pion in V-QCD, a holographic model that incorporates dynamical flavors and chiral symmetry breaking. The model parameters were fixed by earlier fits to hadronic spectra, and no additional parameters were adjusted to form factor data. The resulting functions $\A(t)$ and $\D(t)$ given in (\ref{eq:Atidentification}) and (\ref{eq:Dtidentification}), respectively, thus provide a test of pion structure.

We extracted the matrix element of the energy-momentum tensor (\ref{eq:meldecomposition}) between pion states by applying the LSZ reduction to the holographic three-point function. The external pion states were represented by normalizable solutions of the coupled pseudoscalar and longitudinal axial field equations. The normalization factors entering LSZ reduction were determined from the pion pole residue in the two-point function of the $\bar q \gamma_5 t^a q$ operator. We showed explicitly that the resulting matrix element is independent of the normalization of the pion wave functions. Including both tensor gravity fluctuations and the coupled scalar fluctuations of the metric, dilaton, and the tachyon allowed us to determine $\A(t)$ and $\D(t)$. We showed that the momentum sum rule $\A(0)=1$ follows automatically. Moreover, the terms in $\D(t)$ arising from the spin-zero and spin-two interactions  are separately singular at $t = 0$, but the singularities cancel exactly in the sum. At the physical pion mass, we obtain $\D(0)\approx -0.984$, close to the value $\D(0)=-1$ fixed by the soft-pion theorem in the chiral limit. 

We compared the form factors with results from lattice QCD~\cite{Hackett:2023nkr}, dispersive analyses~\cite{Cao:2024zlf}, Dyson--Schwinger methods~\cite{Xu:2023izo}, and an extraction from pion-pair production data~\cite{Kumano:2017lhr}. Without any parameter fitted to form factor data, our $\A(t)$ is rather close to the lattice data, but its deviation from the result of the dispersive analysis is more significant due to the smaller uncertainty of this analysis. Moreover, our $\D(t)$ decreases in magnitude faster with increasing $-t$ than QCD results indicate. Repeating the computation at the lattice pion mass changes both form factors only slightly, so the difference between our model and the lattice results in $\D(t)$ reflects the model rather than the pion mass. 

From the form factors, we computed the pressure, shear force, and energy density distributions. The pressure is positive at small radii and changes sign at approximately $0.54$ fm, becoming negative at larger radii. The shear force and radial stress are positive for all $\rho$. At the physical pion mass, we obtain a mechanical radius of approximately $0.86$ fm and a mass radius of approximately $1.32$ fm. The energy density integrates to the pion mass but contains a negative region, which calls for caution in interpreting it as a classical mass density \cite{Jaffe:2020ebz}. Although the form factors change little between the two pion masses studied, the radius estimates are more sensitive to the mass because it also enters the Fourier transforms through a relativistic factor, which explicitly depends on $m_\pi$.

The extension of this study to vector and axial-vector mesons is carried out in~\cite{Hippelainen:2026xko}. For these spin-one mesons, the matrix elements of $T^{\mu\nu}$ contain several extra GFFs in addition to the analogs of $\A(t)$ and $\D(t)$. These gravitational form factors provide access to angular momentum and polarization-dependent stress distributions, including quadrupole structure.

Another useful extension is the computation of pion scattering amplitudes using similar holographic methods to those used in the GFF computation. The amplitudes would allow us to extract scattering lengths~\cite{Hoyos:2019kzt,Hoyos:2020fjx} at nonzero quark mass and determine coefficients of the low-energy chiral action~\cite{Sakai:2004cn,Sakai:2005yt,Hoyos:2022ptd}.

\section*{Acknowledgments}
We thank Roman Zwicky for discussions and comments on the draft version of this paper, and Xiong-Hui Cao and Feng-Kun Guo for providing the data from Ref.~\cite{Cao:2024zlf}. N.~J. was supported in part by the Research Council of Finland through grant no. 354533 and the Centre of Excellence in Neutron-Star Physics (project 374062).

\appendix

\section{Choice of potentials in V-QCD}\label{app:potentials}

In this appendix, we give the expressions for the potentials we use in this article. V-QCD potentials have been fitted in the literature to lattice results for QCD thermodynamics~\cite{Gursoy:2009jd,Jokela:2018ers}, meson spectra~\cite{Amorim:2021gat}, and both~\cite{Jarvinen:2022gcc}. Since we are working in the zero-temperature vacuum in this article, we use the fit to meson spectra~\cite{Amorim:2021gat}. To simplify the notation, we introduce
\be
 \lambda = e^\Phi \ .
\ee
The glue potential is\footnote{We have here fixed a misprint appearing in this potential in~\cite{Amorim:2021gat}.} 
\begin{equation}
    \begin{split}
        V_g(\Phi) &= 12 + V_1 \lambda + V_2 \frac{\lambda^2}{1 + \frac{\alpha_\lambda \lambda}{\lambda_0}} + 12 V_\text{IR} e^{-\frac{\lambda_0}{\alpha_\lambda \lambda}} \left(\frac{\alpha_\lambda \lambda}{\lambda_0}\right)^{\frac{4}{3}} \sqrt{\log \qty(1 + \frac{\alpha_\lambda \lambda}{\lambda_0})}  \\
        V_1 &= \frac{44}{9 \pi^2} \ , \qquad V_2 = \frac{4619}{3888 \pi^4} \ , \qquad \lambda_0 = 8 \pi^2 \ ,
\end{split}
\end{equation}
where the values of the near-boundary coefficients $V_1$ and $V_2$ arise from matching the near-boundary flow of $\lambda$ with perturbation theory in the absence of quarks, {\emph{i.e.}}, for improved holographic QCD~\cite{Gursoy:2007cb,Gursoy:2007er}. The flavor potential factors into dilatonic and tachyonic parts, 
\begin{equation}
\begin{split}
        V_f(\Phi, TT^\dagger) &= V_{f_0}(\Phi) V_\tau(TT^\dagger) \ , \qquad V_\tau(TT^\dagger) = (1 + \tilde{a}_1 TT^\dagger) e^{-\tilde{a}_2 TT^\dagger} \\
        V_{f_0}(\Phi) &= W_0 + W_1 \lambda + W_2 \frac{\lambda^2}{1 + \frac{\alpha_\lambda \lambda}{\lambda_0}} + {12 W_\text{IR}} \qty(\frac{\alpha_\lambda \lambda}{\lambda_0})^2 e^{-\frac{\lambda_0}{\alpha_\lambda \lambda}} \qty(1 + \frac{\lambda_0 \widetilde{W}_1}{\alpha_\lambda \lambda})  \\
        W_1 &= \frac{24 + (11 -2x)W_0}{27 \pi^2} \ , \qquad W_2 = \frac{24(857 - 46x) + W_0(4619 - 1714x + 92x^2)}{46656 \pi^4} \ ,
    \end{split}
\end{equation}
where $x = \Nf/\Nc$ is set to $2/3$, and the expressions for $W_1$ and $W_2$ arise from matching the flow of $\lambda$ with the perturbation theory result for full QCD. 
The two remaining potentials to be fixed are
\begin{equation}
    \begin{split}
        \kappa(\Phi) &= 8 \frac{\tilde{a}_2-\tilde{a}_1}{12 - x W_0} \qty[1 + \frac{\alpha_\lambda \kappa_1 \lambda}{\lambda_0} + \Bar{\kappa}_0 \frac{e^{-\frac{\lambda_0}{\alpha_\kappa \lambda}}\qty(1 + \frac{\lambda_0 \Bar{\kappa_1}}{\alpha_\kappa \lambda}) \qty(\frac{\alpha_\kappa \lambda}{\lambda_0})^{4/3}}{\sqrt{\log \qty(1 + \frac{\alpha_\kappa \lambda}{\lambda_0})}}]^{-1}  \\
        w(\Phi) &= w_0 \qty[1 + \frac{\alpha_\lambda w_1 \lambda}{\lambda_0\qty(1 + \frac{\alpha_\lambda \lambda}{\lambda_0})} + \Bar{w}_0 \frac{e^{-\frac{\lambda_0}{\alpha_w \lambda}}\qty(1 + \frac{\lambda_0 \Bar{w_1}}{\alpha_w \lambda}) \qty(\frac{\alpha_w \lambda}{\lambda_0})^{4/3}}{\log \qty(1 + \frac{\alpha_w \lambda}{\lambda_0})}]^{-1} \ ,
    \end{split}
\end{equation}
where the normalization of the kinetic coupling of the tachyon field, $\kappa$, was chosen such that the dimension of the dual operator $\bar q q$ becomes equal to three in the limit $\lambda \to 0$, in agreement with perturbation theory.

\begin{table}[htb]
    \centering
    \begin{tabular}{|c|c|c|c|c|c|}
    \hline
        Parameter & value & Parameter & value & Parameter & value  \\ \hline
        $\alpha_\lambda$ & 2.8328  &  $\alpha_\kappa$  & 3.1670  & $\alpha_w$    &  1.6926 \\ \hline
         $W_0$ & 2.4289 & $\widetilde{W}_1$   & 0.2342  & $W_\text{IR}$   & 1.1345  \\ \hline
         $w_0$ & 0.9400 & $w_1$   & $-0.2898$  &  $\Bar{w}_0$  & 2.9291 \\ \hline
          $\Bar{w}_1$ & 3.0358  & $\kappa_1$   &  1.3254  &  $\Bar{\kappa}_0$  & 1.7647 \\ \hline
         $\Bar{\kappa}_1$ & $-0.3076$ &  $\tilde{a}_1$  & 0.5413 & $\tilde{a}_2$    & 1.5413 \\ \hline
         $V_\text{IR}$ & 1.8042 &  & &  & \\ \hline
    \end{tabular}
    \caption{
    Parameters of the V-QCD potentials at $x = 2/3$, from the fit of~\cite{Amorim:2021gat} to meson spectra.
    }
    \label{tab:actionparameters}
\end{table}

The values of all the additional parameters appearing in the potentials above can be found in Table~\ref{tab:actionparameters}. 

Apart from potentials, to precisely pin down the background geometry and its mapping to QCD, we need to specify the value of the intrinsic scale parameter $\Lambda_{\rm UV}$ and the coefficient $\tau_0$ appearing in the IR asymptotics of the tachyon field (see Appendix~\ref{subapp:solvingthebackground}). We construct two backgrounds. The first is the background used to fit the meson spectrum in~\cite{Amorim:2021gat} at the physical pion mass, $m_\pi \approx 135$~MeV. The second has $m_\pi \approx 170$~MeV. 

The value of $\Lambda_{\rm UV}$ is determined by the near-boundary asymptotics of the geometry and the dilaton field~\cite{Jarvinen:2011qe}. The physical value was not needed and not computed in~\cite{Amorim:2021gat}, because the mass of the $\rho$-meson was used as the reference instead. Setting the $\rho$-meson mass to the physical value, $m_\rho\approx 775$~MeV, sets  $\Lambda_\text{UV} \approx 265$~MeV for the fit of~\cite{Amorim:2021gat}. The value of the coefficient $\tau_0$ in the IR expansion of the tachyon giving this fit\footnote{This value is different from the value quoted in~\cite{Amorim:2021gat} due to a mistake in the asymptotics which does not affect the background or the analysis in this reference.} is $\tau_0 \approx 2.162$. Keeping the potentials fixed and also adjusting the units of energy so that the $\rho$ mass stays fixed, while changing the asymptotics of the tachyon so that the pion mass is $170$~MeV, gives $\Lambda_\text{UV} \approx 261$~MeV at $\tau_0 \approx 2.197$.

\section{Details on solving the V-QCD background}\label{subapp:solvingthebackground}

In this Appendix, we give some details of computing the background solutions numerically for V-QCD.
The background equations \eqref{eq:backgroundeoms} are quite stiff due to the rapid growth of the tachyon in the IR, which makes $V_f$ vanish rapidly there. Here, we describe briefly how these equations are solved. A more complete description of the algorithm with approximate equations for each region can be found in~\cite{Amorim:2021gat}.

First, it is numerically more stable to solve the equations in so-called $A$-coordinates than in $r$-coordinates. In $A$-coordinates, $A(r)$ takes the place of the radial coordinate, and we use
\begin{equation}
    u(A) = \dv[]{r}{A}e^A \ .
\end{equation}
Since $r \simeq \ell e^{-A}$ near the boundary, a uniform grid in $A$ resolves the UV region exponentially finely.

In $r$-coordinates, the relevant IR-asymptotics are
\begin{equation}
    \begin{split}
        A_\text{IR} &= - r^2 + \frac{1}{4}\log(6r^2) - \frac{\log(V_\text{IR})}{2} + \frac{23}{24} - \frac{173}{3456r^2} + O\qty(\frac{1}{r^4})  \\
        \Phi_\text{IR} &= \frac{3}{2}r^2 - \frac{23}{16} - \log \qty(\frac{\alpha_\lambda}{\lambda_0}) - \frac{151}{2304 r^2} + O\qty(\frac{1}{r^4})  \\
        \tau_\text{IR} &\sim \tau_0 r^{\tau_c} \ , \qquad \tau_c = \frac{(12 - xW_0)\Bar{\kappa}_0 \tilde{a}_2\alpha_\kappa^{4/3}}{8 V_\text{IR} (\tilde{a}_2 - \tilde{a}_1) \alpha_\lambda^{4/3}} \ .
    \end{split}
\end{equation}
The background equations of motion are invariant under the combined shift $r \mapsto \Lambda r \ , A \mapsto A - \log (\Lambda)$, that is, with respect to the choice of the unit of energy. We write $A_\text{IR}$ with unit coefficient for $-r^2$, which fixes this freedom.

The only free parameter left is $\tau_0$, which sets the quark mass. We compute two solutions of the background, one where the pion mass is physical, $m_\pi \approx 135$~MeV, and another where $m_\pi \approx 170$~MeV to compare with lattice data.

The background computation proceeds in four stages, in three of which the tachyon does not backreact on the other two fields. This strategy is what reduces the numerical stiffness to make the equations solvable. Between stages, matching is done by requiring the continuity of functions and their derivatives, also providing the necessary initial conditions for the following stage.

In the first two stages we are deep in the IR, where $V_f$ is negligible. There, $u(A)$ and $\Phi(A)$ are solved from the glue part of the equations alone, and the tachyon is then solved in this fixed geometry. In the first stage, it is integrated using the IR-asymptotic form of its equation of motion, with initial conditions from the asymptotics above. The first stage ends at $\tau = 1000/\sqrt{\tilde{a}_2}$.

The second stage is similar to the first one, except that terms suppressed by $1/\tau^2$ in the tachyon equations of motion can no longer be neglected, so one needs to use the full tachyon equation after inserting the separately solved geometry. Other than the updated equation, the tachyon still does not backreact. The second stage ends at $V_f/V_g = V_\text{cut}$, where we have used $V_\text{cut} = 10^{-8}$. 

After this, the third stage requires solving the full coupled equations of motion for all three fields. 

Since $\tau \to 0$ at the UV boundary, its backreaction becomes negligible again close to the UV, where $\tau \propto e^{-A}$ up to power-like corrections. With $A$ normalized by the IR asymptotics, one may safely move to the fourth and final stage for example at $A = 50$, where the tachyon is already exponentially small.

Finally, in the fourth stage the tachyon goes to zero exponentially in $A$ and its backreaction can be dropped, so we solve for $u(A)$ and $\Phi(A)$ separately and then solve a linearized version of the tachyon equation using these profiles. Since working in $A$-coordinates reduces the boundary stiffness of the problem, one may integrate much farther into the UV than in $r$-coordinates. The integration is cut off numerically in the UV and in the IR, for example at $A_\text{UV} = 300$ and $A_\text{IR} = -50$.

\section{Conjugate relations and fluctuation equations}\label{subapp:conjugaterelations}

In this Appendix, we first outline the derivation of the two conjugate relations~\eqref{eq:conjugateequations}, after which we give the fluctuation equations for the tensor mode and coupled scalars, along with a compatibility condition relating the scalar equations to the conjugate relations and a quadratic action from which the scalar equations result.

The conjugate relations~\eqref{eq:conjugateequations} determine $\F$ in terms of $\zeta$ and $\varsigma$ on-shell. Their derivation relies on the linearly perturbed Einstein equations, given in (A.72)--(A.79) of~\cite{Arean:2013tja}, and generalizes the IHQCD treatment of~\cite{Kiritsis:2006ua, Hippelainen:2026izm}.

The first relation is obtained by differentiating the definition of $\F$ in~\eqref{eq:scalarinvariants} and using the $(\mu r)$-component of the perturbed Einstein equations, which gives
\begin{equation}
    \label{eq:firstconjugatestep}
    \F' + 3 A' \F = -\phi + \frac{\psi'}{A'} + \frac{\qty(A')^2 - A''}{\qty(A')^2}\, \psi \ .
\end{equation}
Solving the constraint of~\eqref{eq:gconstraint} for $\phi$ and inserting the background constraint~\eqref{eq:eeconstraint} then gives
\begin{equation}
    \label{eq:firstconjugatederivation}
    \F' + 3 A' \F = \frac{1}{\qty(A')^2}\qty[\frac{4}{9}\qty(\Phi')^2 + \frac{x V_f \kappa \qty(\tau')^2}{6G}]\psi - \frac{4\Phi'}{9A'}\, \chi - \frac{x V_f \kappa \tau'}{6G A'}\, s = \frac{4\qty(\Phi')^2}{9\qty(A')^2}\, \zeta + \frac{x V_f \kappa \qty(\tau')^2}{6G\qty(A')^2}\, \varsigma \ .
\end{equation}

For the second relation in~\eqref{eq:conjugateequations}, one acts with $\Box$ on the definition of $\F$ in~\eqref{eq:scalarinvariants}, solves $\Box\qty(W - E')$ from the dynamical $(\mu\nu)$-trace equation of~\cite{Arean:2013tja}, and eliminates $\phi$ as above. After expressing all derivatives of the potentials through $\pd_\Phi V_f/V_f$, $\pd_\tau V_f/V_f$, and $\pd_\Phi\kappa/\kappa$, using the fact that $\kappa$ and $w$ depend only on $\Phi$, a lengthy but straightforward computation collapses the result onto the second relation in~\eqref{eq:conjugateequations}. Both relations reduce to their IHQCD counterparts~\cite{Kiritsis:2006ua, Hippelainen:2026izm} as $x \to 0$.

We then discuss the fluctuation equations of the dynamical tensor and scalar modes of the geometry, following~\cite{Arean:2013tja}. The tensor mode $\hmntt$ of~\eqref{eq:planewavemodes} is the simplest one, since it does not mix with the scalars. Its propagator obeys the equation of a transverse-traceless graviton in a warped background as in~\cite{DeWolfe:1999cp,Csaki:2000fc,Kiritsis:2006ua, Arean:2013tja}, 
\begin{equation}\label{eq:tensorfluctuationequation}
    \qty(\hmntt)'' + 3 A' \qty(\hmntt)' + \Box \hmntt = 0 \ .
\end{equation}

The equations for coupled scalar invariants $\zeta$ and $\varsigma$ of~\eqref{eq:scalarinvariants} are lengthy, and we write them out explicitly. The two independent scalar invariants satisfy a pair of coupled second-order equations, first derived in~\cite{Arean:2013tja}. In our notation they read \begin{equation}\label{eq:coupledscalarequations}
    \begin{split}
        \zeta'' + k_{\zeta\zeta}\, \zeta' + k_{\zeta\varsigma}\, \varsigma' + \Box \zeta + N_1 \qty(\zeta - \varsigma) &= 0  \\
        \varsigma'' + k_{\varsigma\zeta}\, \zeta' + k_{\varsigma\varsigma}\, \varsigma' + G^2\, \Box \varsigma + N_2 \qty(\varsigma - \zeta) &= 0 \ ,
    \end{split}
\end{equation}
where the coefficients $k_{IJ}$ and $N_{1,2}$ are background functions of $r$. The explicit first-derivative coefficients in~\eqref{eq:coupledscalarequations} are 
\begin{equation}\label{eq:scalarfluctuationcoefficients}
    \begin{split}
        k_{\zeta\zeta} & = \pd_r \log\frac{e^{3A} (\Phi')^2}{(A')^2}= 3A' - \frac{2A''}{A'} + \frac{2\Phi''}{\Phi'} \\
        k_{\varsigma\zeta} &= \Phi'\qty[\frac{4 \Phi' \kappa \qty(\tau')^2}{9 e^{2A} A'} + \frac{1 + G^2}{2}\, \frac{\pd_\Phi \kappa}{\kappa} + G^2\, \frac{\pd_\Phi V_f}{V_f}]  \\
        k_{\zeta\varsigma} &= -\frac{3x}{8}\,\frac{V_f \kappa \qty(\tau')^2}{G^3 \qty(\Phi')^2}\, k_{\varsigma\zeta} =-\frac{3x}{8}\,\frac{V_f \kappa \qty(\tau')^2}{G^3 \Phi'}\qty[\frac{4 \Phi' \kappa \qty(\tau')^2}{9 e^{2A} A'} + \frac{1 + G^2}{2}\, \frac{\pd_\Phi \kappa}{\kappa} + G^2\, \frac{\pd_\Phi V_f}{V_f}]  \\
        k_{\varsigma\varsigma} &=  \pd_r \log\frac{e^{3A} V_f \kappa \qty(\tau')^2}{G^3 (A')^2} \\
        &= \qty(4 G^2 - 7) A' - \frac{2A''}{A'} + \Phi'\qty[\frac{G^2 - 3}{2}\, \frac{\pd_\Phi \kappa}{\kappa} - \qty(2 - G^2)\, \frac{\pd_\Phi V_f}{V_f}] + \frac{2 e^{2A}}{\kappa \tau'}\, \frac{\partial_\tau V_f}{V_f}
        \ ,
    \end{split}
\end{equation}
where all potentials are evaluated on the background. 

The remaining coefficients $N_1$ and $N_2$ are considerably lengthier:\footnote{Note that the expressions for the coefficients given in~\cite{Arean:2013tja} contain misprints which we have corrected here.} 
\begin{equation}\label{eq:N1corrected}
    \begin{split}
        &N_1 \cdot \frac{864}{x}\, \qty(A')^2 e^{4A} G^4 \kappa\, V_f\, \Phi'  = \\
        &-81 \qty(A')^2 e^{4A} G \qty(1 + G^2)^2 \qty(\pd_\Phi \kappa)^2 \Phi' \qty(\tau')^2 V_f^2
        + 288 A' e^{6A} G^5 \kappa\, \Phi' \tau' V_f \qty(\pd_\tau V_f) \\
        &- 324 \qty(A')^2 e^{4A} G^5 \kappa^2 \Phi' \qty(\tau')^2 \qty(\pd_\Phi V_f)^2
        + 324 \qty(A')^2 e^{6A} G^5 \kappa\, \tau' V_f \qty(\pd_\Phi \pd_\tau V_f) \\
        &- 108 A' e^{6A} G^3 \kappa^2 \qty(\tau')^2 V_f^2 \qty(\pd_\Phi V_g) \\
        &- 16 \kappa^2 \Phi' \qty(\tau')^2 V_f^2 \qty{G e^{4A} \qty[36 \qty(A')^2 G^2 \qty(G^2 + 1) - 4 \qty(\Phi')^2 \qty(3 G^2 - 1)] - 3 e^{6A} G^2 \qty(G^2 - 1) V_f x} \\
        &+ \qty(\pd_\Phi V_f) \left\{324 \qty(A')^2 e^{6A} G^5 \kappa\, \tau' \qty(\pd_\tau V_f) \right. \\
        &\left.\qquad + 36 A' e^{2A} G^2 \kappa^2 \qty(\tau')^2 V_f \qty[G e^{2A} \qty(4 \qty(\Phi')^2 \qty(2 - G^2) - 36 \qty(A')^2 G^2) + 3 e^{4A} G^2 V_f x]\right\} \\
        &+ \qty(\pd_\Phi \kappa) \left\{324 \qty(A')^2 e^{6A} G^5 \tau' V_f \qty(\pd_\tau V_f) \right. \\
        &\qquad - 324 \qty(A')^2 e^{4A} G^3 \qty(1 + G^2) \kappa\, \Phi' \qty(\tau')^2 V_f \qty(\pd_\Phi V_f) \\
        &\qquad - 18 A' \kappa \qty(\tau')^2 V_f^2 \left[G e^{4A} \qty(G^2 + 1) \qty(36 \qty(A')^2 G^2 + 4 \qty(\Phi')^2 \qty(G^2 - 2)) \right. \\
        &\left.\left.\qquad\qquad - 3 e^{6A} G^2 \qty(G^2 - 1) V_f x\right]\right\} \ ,
    \end{split}
\end{equation}
\begin{equation}
    \label{eq:N2corrected}
    \begin{split}
        &-N_2 \cdot 2592 \qty(A')^2 e^{6A} G^4 \kappa^2 \tau' V_f^2 = \\
        &+1296 \qty(A')^2 e^{6A} G^4 \qty(1 + G^2) \kappa \qty(\pd_\Phi^2 \kappa) \qty(\Phi')^2 \tau' V_f^2
        - 2304 A' e^{8A} G^4 \kappa \qty(\Phi')^2 V_f \qty(\pd_\tau V_f) \\
        &- 2592 \qty(A')^2 e^{8A} G^6 \kappa\, \Phi' V_f \qty(\pd_\Phi \pd_\tau V_f)
        + 2592 \qty(A')^2 e^{6A} G^6 \kappa^2 \qty(\Phi')^2 \tau' V_f \qty(\pd_\Phi^2 V_f) \\
        &+ \qty(\pd_\Phi V_g) \left\{864 A' e^{8A} G^4 \kappa^2 \Phi' \tau' V_f^2 - 486 \qty(A')^2 e^{8A} G^4 \qty(1 + G^2) \kappa \qty(\pd_\Phi \kappa) \tau' V_f^2 \right. \\
        &\left.\qquad - 972 \qty(A')^2 e^{8A} G^6 \kappa^2 \tau' V_f \qty(\pd_\Phi V_f)\right\} \\
        &+ 324 \qty(A')^2 e^{6A} G^6 \kappa^2 \tau' \qty(\pd_\Phi V_f)^2 \qty(3 e^{2A} G V_f x - 8 \qty(\Phi')^2) \\
        &- 81 \qty(A')^2 e^{4A} G^2 \qty(\pd_\Phi \kappa)^2 \tau' V_f^2 \qty(32 e^{2A} G^2 \qty(\Phi')^2 - 3 e^{4A} G \qty(G^4 - 1) V_f x) \\
        &+ 64 e^{6A} G^2 \kappa^2 \qty(\Phi')^2 \tau' V_f^2 \qty(144 \qty(A')^2 G^2 \qty(2 - G^2) - 16 G^2 \qty(\Phi')^2 - 3 e^{2A} G \qty(G^2 - 1) \qty(3 - G^2) V_f x) \\
        &+ \qty(\pd_\Phi V_f) \left\{2592 \qty(A')^2 e^{8A} G^6 \kappa\, \Phi' \qty(\pd_\tau V_f) \right. \\
        &\left.\qquad - 288 A' e^{4A} G^4 \kappa^2 \Phi' \tau' V_f \qty[e^{2A} \qty(36 \qty(A')^2 G^2 - 4 \qty(\Phi')^2 \qty(2 - G^2)) + 3 e^{4A} G \qty(2 - G^2) V_f x]\right\} \\
        &+ \qty(\pd_\Phi \kappa) \left\{2592 \qty(A')^2 e^{8A} G^4 \Phi' V_f \qty(\pd_\tau V_f) \right. \\
        &\qquad - 144 A' e^{4A} G^4 \kappa\, \Phi' \tau' V_f^2 \left[e^{2A} \qty(3 - G^2) \qty(36 \qty(A')^2 - 4 \qty(\Phi')^2) \right. \\
        &\left.\qquad\qquad - 3 e^{4A} G \qty(G^2 - 1) V_f x\right] \\
        &\left.\qquad + 324 \qty(A')^2 e^{4A} G^4 \kappa\, \tau' V_f \qty(\pd_\Phi V_f) \qty(8 e^{2A} \qty(G^2 - 1) \qty(\Phi')^2 + 3 e^{4A} G^3 V_f x)\right\} \ .
    \end{split}
\end{equation}
In these expressions, the constraint~\eqref{eq:eeconstraint} has been used to eliminate $V_g$, so that the glue potential enters only through its derivative $\pd_\Phi V_g$. 

The coupled system~\eqref{eq:coupledscalarequations} is tied to the conjugate relations~\eqref{eq:conjugateequations} by a nontrivial compatibility condition, which provides a useful consistency check on $N_1$ and $N_2$. Acting with $\Box$ on the first conjugate relation, eliminating $\Box \F$ with the second, reducing $\Box \zeta$ and $\Box \varsigma$ with~\eqref{eq:coupledscalarequations}, and collecting the terms proportional to $\zeta - \varsigma$ yields
\begin{equation}\label{eq:mucompatibility}
    \qty(e^{3A} \I)' + \frac{e^{3A}}{\qty(A')^2} \qty[\frac{4}{9}\qty(\Phi')^2 N_1 - \frac{x V_f \kappa \qty(\tau')^2}{6G^3} N_2] = 0 \ ,
\end{equation}
where $\I(r)$ is the coefficient function of the second conjugate relation in~\eqref{eq:Idef}.
We have checked that the coefficients~\eqref{eq:N1corrected} and~\eqref{eq:N2corrected} satisfy the condition~\eqref{eq:mucompatibility}.

The coupled system~\eqref{eq:coupledscalarequations} also follows from the fluctuation action \begin{equation}\label{eq:scalarquadraticaction}
    \begin{split}
        S^{(2)}_{\zeta\varsigma} = - M^3 \Nc^2 \int \dd^5 x\, \frac{4e^{3A}\qty(\Phi')^2}{9(A')^2} \Big[ &  (\zeta')^2 + \eta^\mn \pd_\mu \zeta\, \pd_\nu \zeta + \frac{3x V_f \kappa (\tau')^2}{8G^3\qty(\Phi')^2}\qty((\varsigma')^2 + G^2 \eta^\mn \pd_\mu \varsigma\, \pd_\nu \varsigma)  \\
        & - 2\, k_{\zeta\varsigma}\, \varsigma' \qty(\zeta - \varsigma) -  N_1 \qty(\zeta - \varsigma)^2 \Big] \ .
    \end{split}
\end{equation}
Varying with respect to $\zeta$ gives the first equation, while varying with respect to $\varsigma$ gives the second, after using the expressions and relations given in~\eqref{eq:scalarfluctuationcoefficients}, and~\eqref{eq:mucompatibility}. In the limit $x \to 0$,~\eqref{eq:scalarquadraticaction} reduces to the Einstein-dilaton action of~\cite{Kiritsis:2006ua}, and its form also agrees with the action derived in~\cite{CruzRojas:2025qcj} for a model with a simpler flavor sector.

\section{The derivation of the linearized pion action}\label{subapp:linearizedpionaction}

Let us expand on the computation of the linearized pion action \eqref{eq:pionaction}. The covariant derivative is expanded to second order in fluctuations as
\begin{equation}\label{eq:covariantderivativeexpansion}
    \D_M T = \pd_M \alphasymbol + i S^1_M + S^2_M \ ,
\end{equation}
with
\begin{equation}\label{eq:S1S2definitions}
    \begin{split}
        S^1_M &= \qty(\pd_M \alphasymbol) \Pi + \alphasymbol \qty(\pd_M \Pi) + 2 \alphasymbol A_M  \\
        S^2_M &= -\frac{1}{2}\qty(\pd_M \alphasymbol) \Pi^2 - \frac{1}{2} \alphasymbol \acomm{\pd_M \Pi}{\Pi} - \alphasymbol \acomm{A_M}{\Pi} + \alphasymbol \comm{V_M}{\Pi} \ ,
    \end{split}
\end{equation}
where we also included a fluctuating vector field $V_M = (L_M+R_M)/2$.
In the symmetrized products entering~\eqref{eq:gaugefieldterms}, the terms linear in $S^{1}$ cancel between the two orderings, and one finds
\begin{equation}\label{eq:symmetrizedproduct}
    \begin{split}
        \qty(\D_M T)^\dagger \qty(\D_N T) + \qty(\D_N T)^\dagger \qty(\D_M T) &= (\D_M T) (\D_N T)^\dagger + (\D_N T) (\D_M T)^\dagger \\
        &= 2 \qty(\pd_M \alphasymbol)\qty(\pd_N \alphasymbol) + 2 \qty(\pd_{\{M} \alphasymbol) S^{2}_{N\}} + S^{1}_{\{M} S^{1}_{N\}} \ .
    \end{split}
\end{equation}
Expanding the products in~\eqref{eq:symmetrizedproduct} and symmetrizing, we find that almost everything cancels: the commutator $\comm{V_M}{\Pi}$ cancels against its conjugate, the terms quadratic in $\Pi$ in $S_M^2$ drop out against terms in $S^{1}_{\{M} S^{1}_{N\}}$, as do the mixed $\qty(\pd \alphasymbol)$-terms. We thus find
\begin{equation}
\begin{split}
    \qty(\D_M T)^\dagger \qty(\D_N T) + \qty(\D_N T)^\dagger \qty(\D_M T) &= 2 (\pd_M \alphasymbol)(\pd_N \alphasymbol) + 2 \alphasymbol^2 (\partial_M\Pi+2A_M) (\partial_N\Pi+2A_N) \ .  
\end{split}
\end{equation}
Collecting the terms of zeroth order in the pion fluctuations into an effective metric $\tilde{g}_\MN$ and the rest into $X_\MN$, we have
\begin{equation}\label{eq:effectivemetricandX}
    \begin{split}
    \boldsymbol{L}_\MN &= \tilde{g}_\MN + X_\MN  \ , \qquad \tilde{g}_\MN = g_\MN + \kappa \qty(\pd_M \alphasymbol)\qty(\pd_N \alphasymbol) \\
    X_\MN &= w F^{L}_\MN + \frac{\kappa}{2} \alphasymbol^2 \acomm{\partial_M\Pi+2A_M}{\partial_N\Pi+2A_N} \ ,
    \end{split}
\end{equation}
and similarly for $\boldsymbol{R}_\MN$. The determinants can be expanded using
\begin{equation}\label{eq:determinantexpansion}
    \begin{split}
        &\sqrt{-\det\qty(\tilde{g}_\MN + X_\MN)} =  \\
        &\sqrt{-\det \tilde{g}}\qty[1 + \frac{1}{2}\tr\qty(\tilde{g}^{-1}X) - \frac{1}{4}\tr\qty(\qty(\tilde{g}^{-1}X)^2) + \frac{1}{8}\qty(\tr\qty(\tilde{g}^{-1}X))^2 + O\qty(X^3)] \ ,
    \end{split}
\end{equation}
where it is understood that the sums over five-dimensional Lorentz indices are implicit.
Note that two different traces appear in the computation: the determinant expansion produces traces over Lorentz indices, denoted by tr, whereas the trace $\Tr$ in~\eqref{eq:VQCDaction} runs over flavor indices and is evaluated only at the end of the computation. 

For the determinant expansion, the three subleading terms of~\eqref{eq:determinantexpansion}  contribute as follows. In the first trace, the field-strength part drops out: the Abelian piece vanishes upon contraction with the inverse effective metric, and the non-Abelian piece is a commutator, removed by the flavor trace, so the first trace contributes only $\frac{\kappa}{2} \tilde{g}^\MN \Omega^2 (\partial_M\Pi+2A_M) (\partial_N\Pi+2A_N)$ (the anticommutator reduces to a product since $\tilde{g}^\MN$ is symmetric). In the second trace only the field-strength terms $\frac{w^2}{4} \qty(F_\MN^{L})^2$ survive, while the last term contributes nothing. 

Since commutator terms are absent, we may combine the left- and right-handed field-strength terms from $\sqrt{-\text{det}\boldsymbol{L}_\MN}$ and $\sqrt{-\text{det}\boldsymbol{R}_\MN}$ as
\begin{equation}\label{eq:LRtoVA}
    \qty(F^{L}_\MN)^2 + \qty(F^{R}_\MN)^2 = 2 \qty(F^V_\MN)^2 + 2 \qty(F^A_\MN)^2 \ ,
\end{equation}
where $F_\MN^V, F_\MN^A$ are the standard Abelian field strengths with $V_M = \frac{1}{2}(L_M + R_M), A_M = \frac{1}{2}(L_M - R_M)$.

Assembling all pieces yields
\begin{equation}\label{eq:quadraticflavoraction}
    \begin{split}
        S_f^{(2)} &= -\frac{1}{2}M^3 \Nc \Tr \int \dd^5 x\, V_f \sqrt{-\det\tilde{g}} \times \Bigg[2 + \kappa \tilde{g}^\MN \alphasymbol^2 (\partial_M\Pi+2A_M) (\partial_N\Pi+2A_N)  \\
        & +\frac{w^2}{2}F^V_\MN \tilde{g}^{MO}\tilde{g}^{NP}F^V_{OP} + \frac{w^2}{2}F^A_\MN \tilde{g}^{MO}\tilde{g}^{NP}F^A_{OP}\Bigg] \ ,
    \end{split}
\end{equation}
where the constant term in brackets reproduces the background flavor action~\eqref{eq:backgroundflavoraction}.

We have written the quadratic flavor action in terms of the vector and axial-vector fields instead of the left- and right-handed fields. The vector field enters only through its field-strength tensor and decouples from the pseudoscalar sector; the same holds for the transverse part of the axial field.  
Both play no further role in the analysis of this article.

\section{Details on the computation of the scalar vertex and the expressions for the form factors}
\label{app:vertexdetails}

Here we collect some of the technical steps behind the analysis of Section~\ref{sec:gravitationalformfactors}, in particular the extraction of the scalar-pion-pion vertex and the properties of the expressions for the gravitational form factors.

\subsection{Reduction of the scalar vertex}\label{subapp:scalarvertexreduction}

We first present the computation leading from the quadratic action~\eqref{eq:pionaction} to the scalar vertex~\eqref{eq:scalarvertexsum}. The strategy is to vary every background quantity in~\eqref{eq:pionaction} with respect to the scalar fluctuations of~\eqref{eq:fluctuatedmetric},
organize the result into a canonical form, and demonstrate that the coefficients of the non-invariant modes $\chi$ and $E$ vanish identically.

The building blocks of~\eqref{eq:pionaction} vary as follows. The effective metric~\eqref{eq:effectivemetricandX} receives the contributions
\begin{equation}\label{eq:effectivemetricvariations}
    \delta \tilde{g}_\mn = 2 e^{2A} \qty(\psi\, \eta_\mn + \pd_\mu \pd_\nu E) \ , \qquad
    \delta \tilde{g}_{\mu r} = e^{2A}\, \pd_\mu W + \kappa \tau' \pd_\mu s \ , \qquad
    \delta \tilde{g}_{rr} = 2 e^{2A}\, \Xi \ ,
\end{equation}
where we have collected the full radial-radial variation into
\begin{equation}\label{eq:Xidefinition}
    \Xi = \phi + e^{-2A} \kappa\, \tau' s' + \frac{e^{-2A} }{2} {\partial_\Phi \kappa}\, \chi \qty(\tau')^2 \ . 
\end{equation}
The measure and the inverse metric vary as
\begin{equation}\label{eq:measureandinversevariations}
    \begin{split}
        \delta \sqrt{-\det \tilde{g}} &= e^{5A} G \qty(4 \psi + \Box E + \frac{\Xi}{G^2}) \ , \qquad
        \delta \tilde{g}^\mn = -2 e^{-2A} \qty(\psi\, \eta^\mn + \pd^\mu \pd^\nu E)  \\
        \delta \tilde{g}^{\mu r} &= -\frac{e^{-2A}}{G^2} \qty(\pd^\mu W + e^{-2A} \kappa \tau' \pd^\mu s) \ , \qquad
        \delta \tilde{g}^{rr} = -2\, \frac{e^{-2A}}{G^4}\, \Xi \ ,
    \end{split}
\end{equation}
and the scalar prefactors as
\begin{equation}\label{eq:prefactorvariations}
    \delta V_f = \qty(\chi\, \frac{\partial_\Phi V_f}{V_f} + s\, \frac{\partial_\tau V_f}{V_f}) V_f \ , \qquad
    \delta \qty(\kappa \alphasymbol^2) = \kappa \tau^2 \qty(\chi\, \frac{\partial_\Phi \kappa}{\kappa} + \frac{2 s}{\tau}) \ , \qquad
    \delta \qty(w^2) = 2 w^2 \chi\, \frac{\partial_\Phi w}{w} \ .
\end{equation}
Finally, the axial field-strength term $F^A_\MN \tilde{g}^{MO}\tilde{g}^{NP}F^A_{OP}$ of~\eqref{eq:pionaction} varies as
\begin{equation}
    \delta\qty(F^A_\MN \tilde{g}^{MO}\tilde{g}^{NP}F^A_{OP}) = - 4 \frac{e^{-4A}}{G^2}\qty(\psi \sfA + \frac{\Xi \sfA}{G^2} + A_\mu'A_\nu' \pd^\mu \pd^\nu E) \ .    
\end{equation}
Carrying out all contractions, we obtain the total variation, with the bilinears $\sfU$, $\sfP$, and $\sfA$ defined in \eqref{eq:bilinears} and $a_i, b_i$ as in~\eqref{eq:aibifactors},
\begin{equation}\label{eq:mastervariation}
    \begin{split}
        \delta S_{A\Pi}^{(2)} & = -\frac{1}{2} M^3 \Nc \Tr \int \dd^5 x
        \left[ \qty(\chi \frac{\partial_\Phi V_f}{V_f} + s \frac{\partial_\tau V_f}{V_f} + 4 \psi + \Box E + \frac{\Xi}{G^2}) \qty(a_1 \sfU + a_2 \sfP + b_2 \sfA) \right. \\
        &\quad + \qty(\chi \frac{\partial_\Phi \kappa}{\kappa} + \frac{2s}{\tau}) \qty(a_1 \sfU + a_2 \sfP) + 2 \chi \frac{\partial_\Phi w}{w}\, b_2 \sfA \\
        &\quad - 2 \Big(a_1 \psi\, \sfU + a_1 \qty(\pd^\mu \pd^\nu E) (\partial_\mu\Pi+2A_\mu) (\partial_\nu\Pi+2A_\nu)   \\
        & \quad + a_2 \qty(\pd^\mu W + e^{-2A} \kappa \tau' \pd^\mu s) \Pi' (\partial_\mu\Pi+2A_\mu) + a_2 \frac{\Xi}{G^2} \sfP\Big) \\
        &\quad \left. - 2 b_2 \qty(\psi\, \sfA + \frac{\Xi}{G^2} \sfA + \sfA_\mn\, \pd^\mu \pd^\nu E) \right] \ .
    \end{split}
\end{equation}
To organize and simplify this expression, we canonicalize: 
all four-dimensional derivatives are integrated by parts such that the scalar fluctuations appear without derivatives (except for $E$, which keeps one derivative, $\pd^\mu E$), the fluctuation equations~\eqref{eq:pioneoms} are used to reduce every pion structure to the bilinears $\sfU, \sfP, \sfA$, and radial derivatives are integrated by parts such that the bilinears appear without derivatives.

Only the $W$- and $s$-dependent term in the third line of~\eqref{eq:mastervariation} takes more than one step: Integrating by parts in $x^\mu$ and using~\eqref{eq:pioneoms}, we obtain
\begin{equation}\label{eq:Wtermcanonicalization}
    \begin{split}
        &-2 a_2 \qty(\pd^\mu W + e^{-2A} \kappa \tau' \pd^\mu s) \Pi' (\partial_\mu\Pi+2A_\mu)  \\
        &\qquad =a_2 \qty(W + e^{-2A} \kappa \tau' s) \qty[\sfU' - \frac{1}{a_1}\qty(2 b_2' \sfA + b_2 \sfA') - \frac{1}{a_1}\qty(2 a_2' \sfP + a_2 \sfP')] \ .
    \end{split}
\end{equation}
We then integrate by parts to remove the the radial derivatives from the bilinears. After these steps, the vertex is a sum of the nine products $\qty{\sfU, \sfP, \sfA} \times \qty{\zeta, \varsigma, \F}$, plus terms proportional to $\chi$ and $E$. As a consistency check, we should show that these additional terms vanish.

Integrating by parts once and expanding the derivatives, the $E$-terms become \
\begin{equation}\label{eq:Etermsintermediate}
    \pd^\mu E \qty[2 a_1 (\partial_\mu\Pi+2A_\mu) \qty(\pd_\nu (\partial^\nu \Pi +2 A^\nu)) - 2 a_2 \Pi' \pd_\mu \Pi' + 2 b_2 \qty(\pd^\nu A'_\nu) A'_\mu] - 2 \qty(\pd^\mu E)'\, a_2 \Pi' (\partial_\mu\Pi+2A_\mu) \ ,
\end{equation}
where the antisymmetric derivative combinations have dropped out, since $\partial_\mu\Pi+2A_\mu$ 
is a gradient. Eliminating $\pd \cdot (\partial \Pi+2A)$ and $\pd \cdot A'$ with the first and third equations in~\eqref{eq:pioneoms}, and writing $\pd_\mu \Pi' = (\partial_\mu\Pi+2A_\mu)' - 2 A_\mu'$, the first term of~\eqref{eq:Etermsintermediate} collapses to a total radial derivative,
\begin{equation}\label{eq:Etermscollapse}
    \pd^\mu E \qty[-2 (\partial_\mu\Pi+2A_\mu) \qty(a_2 \Pi')' - 2 a_2 \Pi' (\partial_\mu\Pi+2A_\mu)'] = -2\, \pd^\mu E \qty(a_2 \Pi' (\partial_\mu\Pi+2A_\mu))' \ ,
\end{equation}
and a final radial integration by parts cancels the last term of~\eqref{eq:Etermsintermediate}.

The $\chi$-terms require more work, since $\chi$ enters nearly every term of~\eqref{eq:mastervariation}. We want to extract only the $\chi$-dependent contributions from each term of the inverted relations~\eqref{eq:invertedscalars} and see that the contributions cancel. For simplicity, with $\varepsilon = \chi / \Phi'$, the inverted fields can be replaced, only for the purposes of this computation, with
\begin{equation}\label{eq:chireplacements}
    \psi \to A' \varepsilon \ , \qquad s \to \tau' \varepsilon \ , \qquad W \to \varepsilon \ , \qquad \phi \to \varepsilon' + A' \varepsilon \ , \qquad \chi \to \Phi' \varepsilon \ ,
\end{equation}
under which $\Xi$ also takes a compact form
\begin{equation}\label{eq:Xichireduction}
    \frac{\Xi}{G^2} = \varepsilon' + \qty(A' + \frac{G'}{G}) \varepsilon \ .
\end{equation}
After canonicalization, the $\chi$-terms organize into the six combinations $\qty{\sfU, \sfP, \sfA} \times \qty{\varepsilon, \varepsilon'}$. The $\varepsilon'$-coefficients cancel immediately. All three of the longer $\varepsilon$-coefficients boil down to the statement that the prefactor combinations reproduce the logarithmic derivatives of the background factors,
\begin{equation}\label{eq:logderivativeidentities}
    \begin{split}
        \frac{a_1'}{a_1} &= \Phi' \qty(\frac{\partial_\Phi V_f}{V_f} + \frac{\partial_\Phi \kappa}{\kappa}) + \tau' \frac{\partial_\tau V_f}{V_f} + \frac{2 \tau'}{\tau} + 3 A' + \frac{G'}{G}  \\
        \frac{a_2'}{a_2} &= \Phi' \qty(\frac{\partial_\Phi V_f}{V_f} + \frac{\partial_\Phi \kappa}{\kappa}) + \tau' \frac{\partial_\tau V_f}{V_f} + \frac{2 \tau'}{\tau} + 3 A' - \frac{G'}{G}  \\
        \frac{b_2'}{b_2} &= \Phi' \qty(\frac{\partial_\Phi V_f}{V_f} + 2 \frac{\partial_\Phi w}{w}) + \tau' \frac{\partial_\tau V_f}{V_f} + A' - \frac{G'}{G} \ ,
    \end{split}
\end{equation}
which hold by the definitions~\eqref{eq:aibifactors}. The coefficient of $\chi$ thus also vanishes identically. This cancelation is the counterpart of energy-momentum conservation and diffeomorphism invariance on the field theory side, and it justifies dropping $\chi$ and $E$ from the computation, which corresponds to the gauge choice $\chi = E = 0$.

With $\chi = E = 0$, the inverted relations~\eqref{eq:invertedscalars} collapse to
\begin{equation}\label{eq:gaugefixedsubstitutions}
    \begin{split}
        \psi &= \zeta \ , \qquad s = \frac{\tau'}{A'} \qty(\zeta - \varsigma) \ , \qquad W = \frac{\zeta}{A'} - \F  \\
        \phi &= \frac{\zeta'}{A'} + \frac{x V_f \kappa \qty(\tau')^2}{6G\qty(A')^2} \qty(\zeta - \varsigma) \ , \qquad \Xi = \phi + e^{-2A} \kappa \tau' s' \ .
    \end{split}
\end{equation}
Substituting~\eqref{eq:gaugefixedsubstitutions} into the canonicalized fluctuation and collecting the coefficients of the nine products $\qty{\sfU, \sfP, \sfA} \times \qty{\zeta, \varsigma, \F}$ then yields the coefficient functions $c_{IJ}$ of~\eqref{eq:scalarvertexcoefficients}. The radial integrations by parts above generate boundary terms, which vanish in the IR by the exponential decay of $V_f$ and hence of every $a_i$ and $b_i$, and in the UV, since pion bilinears are built from the normalizable modes.

\subsection{Regularity of $\D(0)$}\label{subapp:regularityofD}

Let us show that the possible divergence in~\eqref{eq:Dtidentification} cancels analytically, which is equivalent to the condition
\begin{equation}
    \Sigma(0) = \frac{8 m_\pi^2 N_\pi}{M^3 \Nc} \ .
\end{equation}
This requires one to evaluate~\eqref{eq:scalaramplitude} in the limit $t \to 0$. 

In this limit, the scalar fluctuation equations are solved by $\widehat \zeta = \widehat \varsigma = 1$. Using the background Einstein equation~\eqref{eq:eeconstraint}, the first conjugate equation~\eqref{eq:conjugateequations} becomes
\begin{equation}
    \F_0' + 3A'\F_0 = 1 - \frac{A''}{(A')^2} \ ,
\end{equation}
where $\F_0 = \widehat{\mathcal{F}}(0,r)$. To simplify $\Sigma(0)$, we first compute the needed combinations of $c$-coefficients in terms of the above variables. In each combination the $\tau'$- and $c_1$-dependent terms cancel identically. The sums are 
\begin{equation}
    \begin{split}
        c_{\sfU \zeta} + c_{\sfU \varsigma} &= a_2 \qty(1 - \frac{A''}{(A')^2}) + 2 a_1 - \qty(\frac{a_2}{A'})'  \\
        c_{\sfP \zeta} + c_{\sfP \varsigma} &= 4 a_2 - \frac{a_2}{G^2} \qty(1 - \frac{A''}{(A')^2}) + \qty(\frac{a_2}{G^2 A'})' - \frac{2 a_2'}{G^2 A'}  \\
        c_{\sfA \zeta} + c_{\sfA \varsigma} &= 2 b_2 - \frac{b_2}{G^2} \qty(1 - \frac{A''}{(A')^2}) + \qty(\frac{b_2}{G^2 A'})' - \frac{2 b_2'}{G^2 A'}\ ,
    \end{split}
\end{equation}
where we have also used $\frac{a_2(a_1 - a_2)}{a_1} = a_2 - \frac{a_2}{G^2}$ and $\frac{b_2(a_1 - a_2)}{a_1} = b_2 - \frac{b_2}{G^2}$. The remaining coefficients of $\F_0$ are
\begin{equation}
    c_{\sfU \F_0} = a_2' - 3 a_2 A' \ , \qquad c_{\sfP \F_0} = \frac{2 a_2'}{G^2} - \qty(\frac{a_2}{G^2})' + \frac{3 A' a_2}{G^2} \ , \qquad c_{\sfA \F_0} = \frac{2 b_2'}{G^2} - \qty(\frac{b_2}{G^2})' + \frac{3A' b_2}{G^2} \ .
\end{equation}
Now we use the first conjugate equation for the $\F_0$-terms. In particular, we note that, for each $f \in \qty{a_2, -\frac{a_2}{G^2}, -\frac{b_2}{G^2}}$, a term of the form $f' - 3f A'$ appears in the $\F_0$-coefficients. To use the first conjugate equation, multiply it by an arbitrary function $f$ of $r$, so that after rearrangement
\begin{equation}
    (f' - 3A' f)\F_0 = (f\F_0)' - f\qty(1 - \frac{A''}{(A')^2}) \ .
\end{equation}
After using this equation to write each of the $\C$-coefficients, we find that a ``shifted'' field  
\begin{equation}
    Y = \F_0 - \frac{1}{A'} 
\end{equation}
appears in each expression: 
\begin{equation}
    \begin{split}
        \C_\sfU(t = 0) &= 2 a_1 + (a_2 Y)'  \\
        \C_\sfP(t = 0) &= 4 a_2 - \qty(\frac{a_2}{G^2} Y)' + \frac{2 a_2'}{G^2}Y  \\
        \C_\sfA(t = 0) &= 2 b_2 - \qty(\frac{b_2}{G^2} Y)' + \frac{2 b_2'}{G^2} Y \ .
    \end{split}
\end{equation}
Now insert all of these coefficients into the definition of $\Sigma(0)$ to find in total
\begin{equation}
    \begin{split}
    \Sigma(0) = \int \dd r \Bigg\{&-2m_\pi^2 \qty[4 (\psi_P- \psi_L)^2 \qty(2 a_1 + (a_2 Y)') + (\psi_L')^2 \qty(2 b_2 - \qty(\frac{b_2}{G^2} Y)' + \frac{2 b_2'}{G^2} Y) ] \\
    &+ 8 (\psi_P')^2 \qty(4 a_2 - \qty(\frac{a_2}{G^2} Y)' + \frac{2 a_2'}{G^2}Y) \Bigg\} \ ,
    \end{split}
\end{equation}
where we used that $2 p_1 \cdot p_2 = -2 m_\pi^2$ when $t = 0$.

We separate out the terms which include $Y$ and the ones which do not. The latter give
\begin{equation}
    \int \dd r \qty[32 (\psi_P')^2 a_2 - 4 m_\pi^2 \qty(4 a_1 (\psi_P- \psi_L)^2 + b_2 (\psi_L')^2)] \ .
\end{equation}
Using the normalization integrals~\eqref{eq:pionnormalization}, we obtain
\begin{equation}
    \frac{16 M_\pi}{M^3 \Nc} - \frac{8m_\pi^2 N_\pi}{M^3 \Nc} \ .
\end{equation}
By~\eqref{eq:massidentity}, $M_\pi = m_\pi^2 N_\pi$, and the part of $\Sigma(0)$ not dependent on $Y$ gives
\begin{equation}
    \frac{8m_\pi^2 N_\pi}{M^3 \Nc} \ ,
\end{equation}
which is precisely what we needed for regularity. Therefore, the only step left is to show that everything dependent on $Y$ cancels. We collect all terms dependent on $Y$ and integrate by parts the expressions $(fY)'$. The boundary term is 
\be
 (-8m_\pi^2a_2 (\psi_P- \psi_L)^2 +2 m_\pi^2 (\psi_L')^2 \frac{b_2}{G^2} - 8 (\psi_P')^2 \frac{a_2}{G^2})Y \ . 
 \ee 
It vanishes in the IR since $a_2$ and ${b_2}$ carry a factor of $V_f$, vanishing exponentially. It also vanishes in the UV, where $Y \sim r$ up to logarithms, since $A' \sim -1/r$ and, by the first conjugate relation, $\F_0 \sim r$. The coefficients of $Y$ vanish in the UV as well by~\eqref{eq:pionfalloffs}.

We are left with five terms which can be combined into three,
\begin{equation}
    \Sigma(0)_Y = \int \dd r\, Y \qty{16 m_\pi^2 a_2 (\psi_P- \psi_L) (\psi_P'- \psi_L') - \frac{4 m_\pi^2}{G^2} \psi_L' (b_2 \psi_L')' + \frac{16}{G^2} \psi_P' (a_2 \psi_P')'} \ .
\end{equation}
Using the radial equations~\eqref{eq:pionradialeoms}, $(b_2\psi_L')' = -4a_1(\psi_P-\psi_L)$ and $(a_2\psi_P')' = -m_\pi^2 a_1(\psi_P-\psi_L)$, together with $a_1/G^2 = a_2$, the last two terms combine into $16 m_\pi^2 a_2 (\psi_P-\psi_L)(\psi_L'-\psi_P')$, which cancels the first. Thus, $\Sigma(0) = \frac{8m_\pi^2 N_\pi}{M^3 \Nc}$ analytically, and $\D(t)$ is regular.

\subsection{Near-boundary asymptotics and the boundary conditions of gravity fluctuations}
\label{subapp:boundaryasymptotics}

Finally, let us discuss the identity~\eqref{eq:invariantboundaryvalues}. Setting the sources $\chi^{(0)} = s^{(0)} = 0$ does not make the fluctuations $\chi$ and $s$ vanish identically near the boundary, since both retain their VEV contributions, and one must check that these do not contaminate the boundary values of the invariants. The dilaton fluctuation is dual to the $\Delta = 4$ operator $\Tr G_\mn G^\mn$ and the tachyon fluctuation to the $\Delta = 3$ operator $\bar{q} q$, so with the sources turned off, the fields fall off in UV as 
\begin{equation}\label{eq:vevfalloffs}
    \chi \sim r^4 \ , \qquad s \sim r^3 \ ,
\end{equation}
up to logarithmic corrections. The background UV asymptotics of V-QCD~\cite{Jarvinen:2011qe, Arean:2013tja} are 
\begin{equation}\label{eq:backgroundasymptotics}
    A' \sim \frac{1}{r} \ , \qquad \Phi' \sim \frac{1}{r} \ , \qquad \tau' \sim m_q \ ,
\end{equation}
again, up to logarithms. Hence, the VEV contributions entering the invariants behave as
\begin{equation}\label{eq:vevpowercounting}
    \frac{A'}{\Phi'}\, \chi \sim r^4  \to 0 \ , \qquad \frac{A'}{\tau'}\, s \sim \frac{r^2}{m_q} \to 0 \ ,
\end{equation}
as $r \to 0$, \emph{i.e.}, both vanish at the boundary. Hence $\zeta^{(0)} = \varsigma^{(0)} = \psi^{(0)}$, as used in the main text. Note that this power counting relies on $m_q \neq 0$: in the chiral limit the tachyon behaves as $\tau \sim \sigma r^3$, the ratio $A'/\tau'$ grows as $r^{-3}$, and the VEV part of $s$ would contribute at order $r^0$, requiring a more careful treatment. Since the quark mass is finite in our setup, no such subtlety arises.

\section{Large-$t$ extrapolation of the form factors and numerical evaluation of mechanical properties}\label{app:tailfitsandsystematics}

To compute the mechanical properties, we must continue the holographic profiles with pQCD-compatible power-law tails beyond a sufficiently large $-t$. This is needed because the transforms of Section~\ref{sec:mechanicalproperties} extend to $- t \to \infty$, while our numerical calculation stops at finite $-t$. We continue the numerical profiles with tails of the power-law form
\begin{equation}\label{eq:tailAnsatz}
    \A(t) \sim \sum_{i = 1}^N \frac{\A_i}{(-t)^i} \ , \qquad N = 3 \tx{or} 4 \ ,
\end{equation}
and analogously for $\D(t)$, with $\A_i$ and $\D_i$ in units of GeV$^{2i}$. The coefficients are determined by least-squares fits to our numerical profiles over windows at the large-$|t|$ end of the data, and the integrals are computed in two pieces, numerically up to the matching point and analytically over the tail.

\begin{figure}[!htb]
    \centering
    \includegraphics[width=0.8\linewidth]{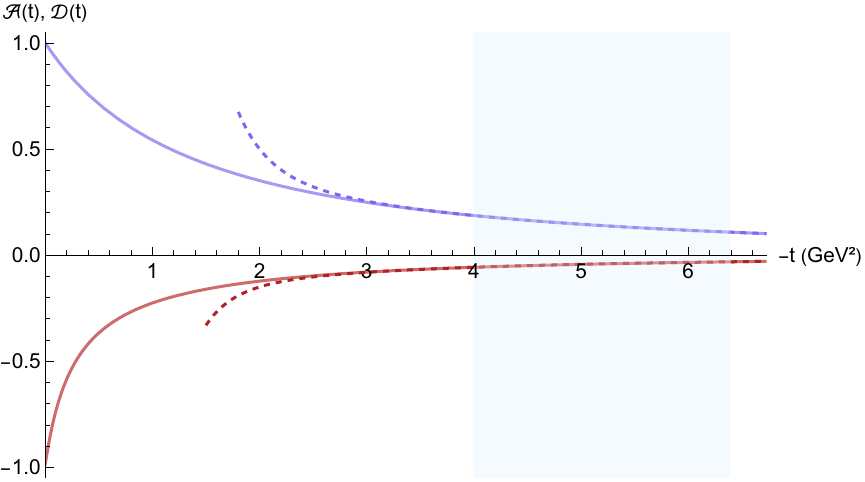}
    \caption{
    Our holographic form factors $\A(t)$ and $\D(t)$ at $m_\pi \approx 135$~MeV (solid curves) and the corresponding four-term power-law fits (dashed curves). The shaded band indicates the fitting window $-t \in [4.0, 6.4]$~GeV$^2$.}
    \label{fig:formfactorfits}
\end{figure}

We perform four different fits for both form factors, either with three or four terms, and either fitting data in the window $-t \in [4.0, 6.4]$~GeV$^2$ or $-t \in [3.0, 5.4]$~GeV$^2$. An example fit with four terms and the higher window is shown in Fig.~\ref{fig:formfactorfits}.

\begin{figure}[!htb]
    \centering
    \includegraphics[width=0.8\textwidth]{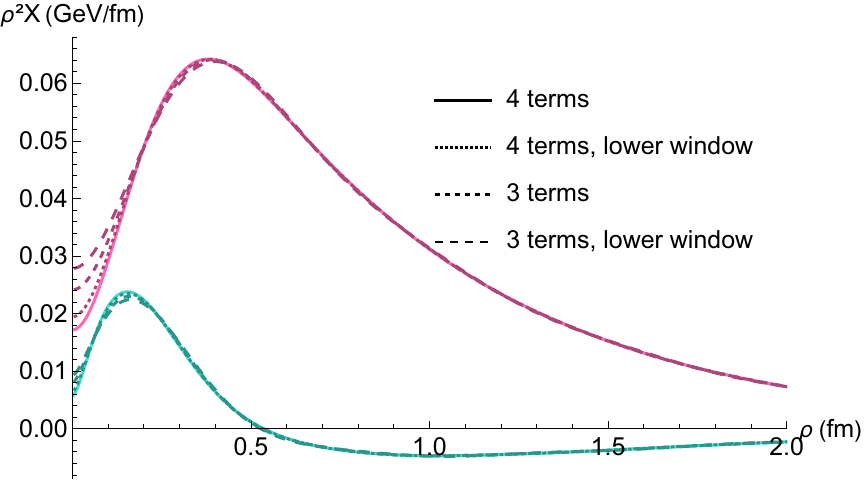}
    \caption{Mechanical profiles obtained using different asymptotic fits. The turquoise curves show $\rho^2 p(\rho)$, and the pink curves show $\rho^2 s(\rho)$.}
    \label{fig:tailsystematics}
\end{figure}

\begin{table}[htb]
    \centering
    \begin{tabular}{|c||c|c|c|c|c|c|c|c|}
    \hline
        & $\A_1$ & $\A_2$ & $\A_3$ & $\A_4$ & $\D_1$ & $\D_2$ & $\D_3$ & $\D_4$  \\
    \hline
    \hline
        4 terms & 0.273 & 4.67 & -15.4 & 17.3 & -0.071 & -1.30 & 3.62 & -3.76  \\ \hline
        4 terms, lower window & 0.360 & 3.50 & -10.2 & 9.52 & -0.088 & -1.08 & 2.63 & -2.29 \\ \hline
        3 terms & 0.414 & 2.55 & -4.88 & -- & -0.102 & -0.841 & 1.33 & -- \\ \hline
        3 terms, lower window & 0.517 & 1.61 & -2.79 & -- & -0.126 & -0.627 & 0.855 & -- \\
         \hline
    \end{tabular}
    \caption{Tail fit parameters of~\eqref{eq:tailAnsatz}, in units of GeV$^{2i}$.
    }
    \label{tab:tailfitparameters}
\end{table}

As a numerical robustness check, we compute the pressure and shear profiles using all four fits and compare them in Fig.~\ref{fig:tailsystematics}. Changes to the asymptotic tails primarily affect the profiles at small $\rho$. The results are similar across the four fits. In the main text, we use the four-term fit over $-t\in [4.0,6.4]\,\mathrm{GeV}^2$. The parameters of all four fits are listed in Table~\ref{tab:tailfitparameters}.

\bibliographystyle{JHEP}
\bibliography{bibliography.bib}

\end{document}